\documentclass[acmsmall, authorversion, nonacm]{acmart}

\AtBeginDocument{%
  \providecommand\BibTeX{{%
    \normalfont B\kern-0.5em{\scshape i\kern-0.25em b}\kern-0.8em\TeX}}}

\renewcommand\footnotetextcopyrightpermission[1]{} % removes footnote with conference information in first column
\acmJournal{TCPS}

\setcitestyle{numbers,sort&compress}

\usepackage{xcolor}
\usepackage{subfig,graphicx}
\usepackage{enumitem}
\usepackage{listings}
\usepackage{multirow}
\usepackage{graphicx}
\usepackage{verbatim}
\usepackage{url}
\usepackage{mathtools}
\usepackage{adjustbox}
\usepackage{caption}
\usepackage[frozencache]{minted}
\usepackage{array}
\newcolumntype{Q}{>{\centering\arraybackslash}m{2.8cm}}
\newcolumntype{D}{>{\centering\arraybackslash}m{1cm}}

\begin{document}
\title{Flexible, Decentralized Access Control for Smart Buildings with Smart Contracts}

\author{Leepakshi Bindra}
\email{leepaksh@ualberta.ca}
\author{Kalvin Eng}
\email{kalvin.eng@ualberta.ca}
\author{Omid Ardakanian}
\email{ardakanian@ualberta.ca}
\author{Eleni Stroulia}
\affiliation{%
\institution{University of Alberta, Canada}
\city{Edmonton}
\country{Canada}
}
\email{stroulia@ualberta.ca}

\renewcommand{\shortauthors}{Bindra, et al.}

\begin{abstract}
Large commercial buildings are complex cyber-physical systems containing expensive and critical equipment that ensure the safety and comfort of their numerous occupants. Yet occupant and visitor access to spaces and equipment within these buildings are still managed through unsystematic, inefficient, and human-intensive processes. As a standard practice, long-term building occupants are given access privileges to rooms and equipment based on their organizational roles, while visitors have to be escorted by their hosts. This approach is conservative and inflexible.
In this paper, we describe a methodology that can flexibly and securely manage building access privileges for long-term occupants and short-term visitors alike, taking into account the risk associated with accessing each space within the building. Our methodology relies on blockchain smart contracts to describe, grant, audit, and revoke fine-grained permissions for building occupants and visitors, in a decentralized fashion. 
The smart contracts are specified through a process that leverages the information 
compiled from Brick and BOT models of the building. 
We illustrate the proposed method through a typical application scenario in the context of a real office building 
and argue that it can greatly reduce the administration overhead, while, at the same time, 
providing fine-grained, auditable access control.
\end{abstract}




\maketitle

\thispagestyle{empty}

%-----------------------------------------
\section{Introduction}\label{sec:intro}
%-----------------------------------------

%ES: Points to make:
%1) we need a single coherent mechanism for managing access to a building's spaces and systems
%2) the mechanism has to be flexible since any commercial building is bound to be the home of many organizations, with different roles and policies 
%3) the mechanism has to be easy and robust: distributed and easy to manage
%4) the process has to be auditable

Modern commercial buildings are complex cyber-physical systems. They are increasingly being equipped with sensors and actuators, ranging from surveillance cameras and card readers for security and access control, to thermostats and air-quality sensors feeding into the Heating, Ventilation, and Air Conditioning (HVAC) system, which controls the indoor environment while maintaining occupants comfort~\cite{Haggerty13,ardakanian2018}. These buildings represent substantial financial investments and the management of their security is even more critical compared to older buildings.

Reasoning about and managing access to these buildings require different access policies for different types of users. For example, as a standard practice, long-term building occupants such as employees who work in the building are given access privileges to their offices and shared spaces, based on their organizational roles; on the other hand, facilities-management personnel typically have access to the more restricted spaces where equipment is installed and also to the Building Management System (BMS) which enables them to monitor and control the equipment settings. Occupants may access the sensing and control devices, such as light switches and thermostats, in the spaces to which they have physical access even though these devices may impact building areas beyond the room in which they are physically located; for example, a thermostat located in a room can determine the temperature setpoint of multiple adjacent rooms. Finally, visitors tend to have limited access, and are frequently required to be escorted by building occupants to the meeting rooms where their business is taking place; they might control equipment in these rooms, but only for the duration of their meeting. In large commercial buildings, this approach implies substantial administrative overhead and exposes the building infrastructure to various security threats.
%since physically access to a room implies access to equipment and corresponding sensing and control points.
%Commercial buildings are difficult to navigate for any new person, and can be overwhelming if the person has to reach a particular room in the least amount of time. For instance, Bob is a new employee in an organization and has to reach a meeting room in his new office building. In order to manage his time well, he would like to know the way from the entrance of the building to the meeting room. He may also need to pass doors in the hallway that require every employee to scan their card and thus, would require access to such spaces a prior. Along with these, he should also be careful of not entering sensitive and restricted spaces.

Ideally, an automated solution is needed to efficiently manage the access privileges of a building's occupants. 
Traditional role-based access-control models adopted in existing access control systems are unwieldy, 
in that they require the specification of all roles and their relative authority, 
which is a challenge in large buildings occupied by multiple organizations, 
each one with their own different role hierarchy~\cite{Bindra19}. 
In this paper, we propose a methodology for managing building access control inspired by the concept of \emph{airline boarding-passes} and the workflows around them. At any point in time, the airport is used by numerous airlines that manage their own flights at their corresponding gates and 
are responsible for issuing boarding passes to their passengers. 
Boarding passes become available shortly before passengers travel, enabling passengers to go through security, access their gate, and board their plane at the right time. Each boarding pass is associated with an individual traveller, and is valid only during a short period before the flight departure. During this period, security personnel are able to scan and verify the boarding pass. This methodology is envisioned to be implemented independent of any pre-existing access control delegation strategies, but still be co-existent with already implemented access control systems.
%The SmartBuildingPass will use blockchain to issue smart-contract enabled passes to building visitors. The moment an appointment is set for a visitor to a Smart Building, the visitor will receive an electronic pass that will let them access the meeting room and all the passageways leading from the main entrance to the specific meeting room, within the relevant time period. The pass may be implemented with badges/bracelets for the long-term occupants of the building and with a mobile app for visitors. Privileges will be flexibly managed based on the BIM model of the building, and the underlying smart contracts will be appropriately maintained. 

Our methodology can flexibly and securely manage building access privileges for long-term occupants and short-term visitors alike, 
addressing the challenges and risks mentioned above. We establish a set of services for 
{\em fine-grained} decentralized management of people's access privileges within a commercial building. We use the term fine-grained to mean
(a) person-centric instead of role-based, 
(b) tailored to different space/system granularities, and 
(c) spanning multiple timescales. 
The underlying intuition for this work is that 
{\em if a person is authorized to have physical access to a particular location in a building, 
then they also have access and opportunity to manipulate 
the sensors and control points in this space.}
This is because, in most cases, there is no additional access-control 
beyond placing the equipment behind a locked door.
Also by implication, {\em if a person should not have access to some control points, 
they should not be authorized to access the space where this equipment is located.} 

Our methodology relies on blockchain smart contracts to describe, grant, audit, and revoke fine-grained permissions for building occupants and visitors in a decentralized fashion~\cite{andersen2018democratizing}. The smart contracts are specified through a process that leverages the information compiled from the BOT~\cite{Rasmussen17} and Brick~\cite{balaji2016brick} models of the building's spatial structure, equipment, and their relations. This information enables our methodology to grant an individual with just the right access privileges to let them reach their destination within the building. We implement this intuition in the form of smart contracts, through which space-access privileges are given to (and revoked from) individuals using an API that can be invoked by different software applications. The underlying \emph{access-control service} is responsible for accepting or rejecting individual access requests based on the currently valid smart contracts and their implications regarding access to sensors and actuators. 
We illustrate the proposed method through a typical use case in the context of a real building and we argue that it can greatly reduce the administration overhead, while at the same time providing fine-grained, auditable access control.

The rest of this paper is organized as follows. 
Section~\ref{sec:background} provides an overview of the context around our work.
Section~\ref{sec:method} describes in detail our methodology for smart-building access control using a graph model for reasoning, and blockchain smart contracts for managing access.
Section~\ref{sec:implement} examines the implementation of the proposed work along with use cases in which the proposed solution can greatly enhance the authorization and access control procedures in a real building.
Section~\ref{sec:perform} evaluates the performance of the proposed solution in terms of throughput and delay.
Section~\ref{sec:system} presents practical considerations for using blockchain-based access control.
Section~\ref{sec:related} discusses related work on authorization and access-control solutions developed for the built environment. 
The paper concludes with an overview of the key contributions of our work and our plans for future work in Section~\ref{sec:conclusion}.
%-----------------------------------------
\section{Background}\label{sec:background}
%-----------------------------------------
We review key concepts and technologies on which our method relies in this section. 
Access-control privileges are managed through smart contracts implemented on a private Ethereum blockchain~\cite{buterin2014}. 
The reasoning process producing the smart contract as a sequence of doors, spaces and devices accessible to the visitor relies on the building's information captured in the BOT and Brick models. %These two ontologies are described below.

%==========================================
\subsection{Access-Control Paradigms}
%==========================================
Access control regulates what resources users may use, based on their assigned privileges. In principle, there are three general mechanisms for reasoning about what permissions should be given to a user. Extensions to these paradigms include risk-aware access control which associates a cost to providing access, and access control using
building information models (BIMs) which incorporates the knowledge of building into access control.

{\em Role-based} access control relies on an explicit, and fairly static, list of organization roles associated with privileges. In this model, each user is associated with a role, which entitles them to a set of privileges corresponding to their role(s)~\cite{sandhu1996role}. Commercial buildings usually host multiple organizations, each of which defines its own role hierarchy which is relevant to the areas of the building that it occupies; as a result, there is no single role hierarchy that pertains to the building as a whole. Furthermore, often times there is no single central authority who can manage roles for all building spaces. 

{\em Risk-based} access control is a model where users are granted access to resources based on a scoring function that dynamically and contextually quantifies risk implicit in this privilege~\cite{kandala2011attribute}. This approach to access control is more relevant in dynamic environments, where the specific context of the access request should inform whether the request may be honored or not. Buildings are not that dynamic and, in principle, a more explicit, less contextual, access-control mechanism is desirable.

{\em Attribute-based} access control grants access rights to users through the use of policies that combine together (with logical operators) different user, resource, object, and environment attributes~\cite{wang2004logic}. In our work, we adopt this paradigm to develop a {\em cost function} that represents the sensitivity of building spaces based on their function and equipment they contain. A room with many control points, occupied by an employee in a position of authority in the organization, is more sensitive (and is, therefore, associated with a higher cost) than the building's reception for example. In principle, this cost function enables access-control policies to be defined based on sensitivity ranges. It also enables one to reason about the relative sensitivity of spaces and rationalize the access-granting process.
%In non-risk based access control models, there is no measurable basis for decisions that are made when granting access.

\subsubsection{Risk-Aware Access Control}
In risk-aware access control, a risk (defined by a cost function) 
is associated to each user who wishes to access resources.
It is then compared against a predetermined threshold, before the user is granted access.
This differs from traditional access control models that have predefined policies set for granting access 
and can be more permissive.
The problem arises when ``low-risk'' users are automatically granted access 
to resources that were never intended to be accessed by them.
In this work, instead of associating risks to users,
we adopt a cost function to help develop access control policies 
where the acceptable risk level can be defined a priori.

Several cost functions have been proposed for access control. 
Chen et al.~\cite{chen2011risk} propose cost functions for defining risk that incorporate 
the trustworthiness of a user, the degree of competence of a user with respect to a particular user-role assignment, 
and the degree of appropriateness of a permission-role assignment for a given role. 
Salim et al.~\cite{salim2013budget} consider the monetary value of a resource or an inferred impact of misusing it
(when monetary value is unavailable) as a basis for their cost function. 
Bijon et al.~\cite{bijon2013framework} propose a risk-based access control framework 
that incorporates the quantified risk for granting access and specific thresholds
calculated based on attributes, purpose, and situational factors.
Inspired by these cost functions, 
we take semantics and relationships of spaces and resources into account
when defining the amount of risk associated with accessing spaces and resources therein.

\subsubsection{BIM-based Access Control}
Access control policies can be developed leveraging BIM.
Skandhakumar et al.~\cite{skandhakumar2012authorization} provide a review of 
spatio-temporal access control models and 
propose an authorization framework that involves (a) modeling of spatial data in BIM, 
(b) creation of access policies based on BIM, and (c) authorization of these policies. 
In particular, the authors introduce `contains', `connected', `adjacent', and `accessible' 
relationships between building elements which are accounted for when reasoning about access policies. 
% The framework introduced is implemented in a tool that allows for access decisions to different building locations 
% using a 3D model for better spatial reasoning to assign/revoke and audit/monitor access~\cite{Skandhakumar2012BIM}. 
% The tool's priority is to delegate access to locations on a least-privilege basis, i.e.,
% people who are granted access are granted permission to the spaces 
% that provide the least amount of access to other spaces directly or indirectly. 
To capture the relationships between spaces in a building, 
BIM is transferred to a graph model in~\cite{skandhakumar2016graph}. 
% The graph model, to the best of our knowledge, is the first formalized instance of building modeling 
% that predates BOT modeling of building spaces~\cite{skandhakumar2016graph}. 
Despite the novelty of this model, it does not incorporate concepts such as sensors, actuators,
and building subsystems which can be affected by people who are given access to the building spaces.
To specify access control policies, 
the use of `eXtensible access control markup language' (XACML) is proposed in~\cite{skandhakumar2018policy}. 
XACML is a standard language for specifying and evaluating access requests. 
Our proposed smart contract solution is similar to XACML in that 
we separate authorization across different services and 
provide a template smart contract to execute access requests.

% \textcolor{blue}{
% The shortcomings of using BIM are that it mainly captures structural information of how a building is built but fails to capture the inherent relationships between resources and subsystems within a building. 
% Furthermore, the use of a non-standardized graph model presents 
% an integration issue when trying to incorporate resources and subsystems. 
% In order to alleviate the problem of BIM, we use the Brick~\cite{balaji2016brick} ontology 
% to capture the inherent relationships between resources and subsystems, and align it with the BOT ontology to capture spatial relationships. As a result of aligning the Brick and Bot ontologies together, it allows for a simple platform to perform SPARQL graph queries to understand the building and calculate optimal pathways, and interoperability that can be extended to more buildings, resources, and subsystems.
% }

%==========================================
\subsection{Blockchain and Smart Contracts}
%==========================================
Blockchain is a distributed and shared ledger that serves as an irreversible and incorruptible public repository~\cite{narayanan2016}. It enables the occurrence of a particular transaction without requiring a central authority. %to manage all the transactions. 
Compared to traditional database systems, it offers three major advantages: 
\begin{itemize}
\item As a {\em distributed} system, blockchain eliminates the need for a costly infrastructure that relies on prox cards for occupants and requires one of these occupants (or dedicated security personnel) to escort visitors to their meeting locations using their own prox cards. %and mitigates the risk of tampering. %and promises increased security by eliminating ``middle-man'' attacks.
\item Blockchain does not require a trusted third party to certify transactions, thanks to public-key cryptography and a consensus mechanism. This allows digital transactions to occur between parties that do not have pre-established trust relations, i.e., {\em trustlessness}.
\item The state stored in blockchain is {\em immutable} due to the use of cryptographic hash functions.
\end{itemize}

A smart contract is executable code that runs on a blockchain to facilitate, execute, and enforce the terms of an agreement between different parties. It can also be used to encode an arbitrary state-transition function.
% It can be thought of as a system that automatically moves digital assets to all, or some, of the involved parties 
% according to the rules specified in the contract~\cite{buterin2014next}. 
%These do not require a trusted third party to approve the transaction. Once deployed on the blockchain, the contract is immutable.
Each contract is assigned a unique address. 
Users can send a transaction to this address for execution.
A callback function is executed when a transaction execution request is received. 
If the transaction is successfully completed, the contract's state is updated. 
Otherwise, any changes made to the state are reverted. %and the transaction is unsuccessful.

The development of smart contracts is supported by many blockchain platforms, 
Ethereum~\cite{buterin2014} being the most well known and, perhaps, the most broadly adopted one. Ethereum provides an abstract foundation layer for smart-contract development: a blockchain with a built-in Turing-complete programming language for the specification of smart contracts, with arbitrary rules for ownership, transaction formats and state-transition functions. Ethereum has its own cryptocurrency called \emph{ether} and an internal currency to pay for computations and transaction fees called \emph{gas}.

%The code in Ethereum contracts is written in a low-level, stack-based bytecode language, referred to as "Ethereum virtual machine code" or "EVM code". The code consists of a series of bytes, where each byte represents an operation. In general, code execution is an infinite loop that consists of repeatedly carrying out the operation at the current program counter (which begins at zero) and then incrementing the program counter by one, until the end of the code is reached or an error or STOP or RETURN instruction is detected\cite{buterin2014next}.

Drawing on~\cite{Andersen:EECS-2017-234} that argues for democratized access to the physical resources in buildings using a blockchain, 
we utilize a private Ethereum blockchain to store the \emph{authorization graph} of a specific building. 
Executing a transaction (e.g., for adding or revoking users' accesses) leads to a state change and updates this graph. 
The delegation of trust can be performed by any user without communicating with a central authority. 
The state can be read from the blockchain to verify access for any user at any time. 
% \textcolor{red}{We use a private blockchain as a proof-of-concept implementation, 
% but our method can be extended to public and hybrid blockchains 
% after considerations highlighted in Section~\ref{sec:system}.}

Our proof-of-concept implementation uses a private Ethereum blockchain 
over a public blockchain for performance and privacy reasons\footnote{Private blockchains 
can achieve better scalability than public blockchains, thanks to the limited number of participants they have,
and can minimize privacy concerns since only authorized users are allowed to connect and perform transactions. 
Furthermore, the transaction cost can be less of a concern as it is not tied to the volatile cryptocurrency market.}. 
The private blockchain network is comprised of several nodes 
representing different groups within an organization 
or different organizations housed in the same commercial building. 
Compared to a centralized access control system which runs a private server,
it provides better transparency, availability, and robustness;
there is no single trusted entity and no single point of failure,
the integrity of access-related data is always maintained and 
tenant organizations can easily audit transactions.
Although privacy is easier to achieve in centralized systems, 
the private blockchain helps preserve privacy to some extent 
as only specific nodes within the organization are allowed to connect to the network.
We note that private permissioned blockchains, 
such as the Hyperledger Fabric, and hybrid blockchains
could provide advantages similar to a private Ethereum blockchain
and can be considered as alternative solutions.
%The use of private blockchain has been proposed in previous work for controlling access to building subsystems in a decentralized and democratized fashion \cite{andersen2019wave} \cite{andersen2018democratizing}. In this work, we extended this idea to controlling access to building spaces in addition to sensing and control points therein.

% \textcolor{red}{Blockchains can be public, private, or a hybrid.
% Private blockchains, such as Hyperledger Fabric, 
% can achieve better scalability than public blockchains thanks to the limited number of participants they have,
% and minimize privacy concerns as only authorized users are allowed to connect and perform transactions. 
% Furthermore, transaction cost can be less of a concern as it is not tied to the volatile cryptocurrency market.
% As opposed to public blockchains, where smart contracts will likely have to be obfuscated to maintain privacy, and transactions can vary greatly as its value is dependent on cryptocurrencies. 
% However, an advantage of public blockchain is that it allows for greater resilience of transactions. 
% Hybrid blockchains attempt to take the best of private and public blockchains 
% to address some of the shortcomings of private and public blockchains.}

%==========================================
\subsection{Building Models}
%==========================================
%ES: discuss tag/relation inconsistencies
%KE: talk about the bot ontology too that describes spatial relations?
%KE: what about a ontology for sensitivity levels?

The lack of a common representation for buildings has historically hindered the development of portable building applications. To address this issue, several standards for modelling building data have been conceived in recent years, examples of which are Project Haystack~\cite{haystack} and Brick~\cite{balaji2016brick}. The Brick schema~\cite{balaji2016brick} defines an ontology for describing the various building spaces and subsystems, their components, and relationships between them. It defines three types of {\em entities}: locations, equipment, and points. Locations are hierarchically organized, in terms of buildings, floors, and rooms. Equipment may be composed of many parts and may be connected to other equipment with certain functional relationships. They comprise complex building subsystems, such as HVAC, lighting, and plumbing. Sensors and setpoints are two types of physical points that can generate timeseries data and are used in control loops of different equipment.

Brick describes a building through a collection of {\em triples} ({\em subject - predicate - object}), following the Resource Description Framework (RDF) data model. Each triple consists of two entities connected with a relationship, which can be \textsc{feeds}, \textsc{controls}, \textsc{hasPart}, \textsc{hasPoint}, or \textsc{isLocationOf}. The collection of such triples forms a directed graph, where nodes represent the entities and edges represent the relationships between them. Figure~\ref{fig:infer} shows a subset of Brick entities and their relationships in an example building. The RDF syntax allows for using SPARQL (the RDF query language) to reason about various entities and relationships. For example, it is possible to retrieve sensors and control points that are located in a specific room or floor of a building, and are used to control the operation of a given Variable Air Volume (VAV) system.

Despite the effort to make Brick expressive enough to capture all important relationships between different building entities, it does not capture the adjacency relationship between different spaces in a building. This information is essential for building applications such as indoor path planning. The Building Information Model (BIM)~\cite{SUCCAR2009357}, which supports computational methods for designing and constructing buildings, captures information about the building interior layout and adjacency of rooms, %(typically found in 3D architectural drawings), 
but unlike Brick, it lacks mechanisms for describing functional relationships between different entities~\cite{balaji2016brick}. BIM has been traditionally represented using the Industry Foundation Classes (IFC)~\cite{bazjanac1999industry} data model designed to facilitate interoperability in the building industry. BIM exhaustively describes composition of building subsystems, it leads to unnecessary complexity when capturing information about a building. This has motivated the Linked Building Data (LBD) Community Group\footnote{\url{https://www.w3.org/community/lbd/}} to create the Building Topology Ontology (BOT)~\cite{Rasmussen17}. BOT is a minimal ontology for describing the spatial structure of a building. It defines three types of {\em entities}, namely zones, elements, and interfaces, and captures relationships, such as {\em adjacency} and {\em containment} between these entities. Zones are hierarchically organized in terms of sites, buildings, storeys, and spaces, which are spatial 3D divisions used to model rooms. Elements are physical building components such as doors and walls. An interface is the surface where two elements, two zones, or an element and a zone meet. A subset of BOT entities and their relationships are depicted in Figure~\ref{fig:infer} for an example building. Aligning BOT with Brick allows for creating a unified RDF model of building's structure and equipment, thereby enabling the use of SPARQL to reason about the sequence  of doors and rooms that need to be traversed to go from one location to another location in a building, and equipment that can be accessed in these rooms.

\section{Related Work}\label{sec:related}
%-----------------------------------------
Conventional methods for authentication, authorization, and revocation rely on a trusted central authority. 
For example, the existing authorization method LDAP~\cite{zeilenga2006lightweight} 
uses Role-Based Access Control (RBAC) with a single central authority. 
Kerberos~\cite{neuman1994kerberos} and Jabber~\cite{saint2005streaming} are similar in that respect. 
Several systems are also developed to eliminate the central authority, 
examples of which are CCN~\cite{jacobson2009networking} and the Web~of~Trust~\cite{callas2007openpgp,caronni2000walking}. 
They adopt a decentralized peer-to-peer trust model in which a principal, denoted by a public key, 
can publish a signature of another public key to denote trust. 
SmartTokens~\cite{dmitrienko2012smarttokens} relies on a token-based access control system 
for NFC-enabled smartphones, in which the delegation of access to other smartphone users 
can be accomplished without a central authority. 
Although SmartTokens uses symmetric cryptography, 
users need to present all delegated tokens through the delegation chain in order to be verified.
In recent work~\cite{le2019access}, a lightweight distributed authorization protocol is proposed
supporting delegation of access right to a smart device in the form of a Bloom filter. 
This method of delegation uses secured hashing to prevent the permission from being forged. 
%However, it claims to have only sufficient security due to the false positive rates of Bloom filters.

RBAC-SC~\cite{cruz2018rbac} performs role-based access control using blockchain 
and a challenge-response protocol for authentication.
FairAccess is a cryptocurrency blockchain-based access control framework~\cite{ouaddah2016fairaccess}.
This work is different from ours as a distinct smart contract is created
for the access control policy of every resource-requester pair. 
Furthermore, they include the IoT devices in the blockchain, 
whereas there are IoT devices that do not have the capabilities to run the blockchain on them.
In another line of work, a distributed architecture called ControlChain is proposed~\cite{pinno2017controlchain}. 
ControlChain enables the expression of a wide variety of access control models, 
such as RBAC~\cite{ferraiolo1995role}, OrBAC~\cite{kalam2003organization} and ABAC~\cite{hu2013guide}, deployed on IoT. 
This mapping is enabled due to the use of a \emph{Decoder} entity 
that automates the translation of access control model and rules to their supported mechanisms. 
However, none of the these decentralized systems
addresses how building metadata can be linked with smart contracts 
to extend access control to spaces and equipment in the building.

\subsection{Access Control in Buildings}
%===================================================
Using smart contracts for authorization and delegation of trust was originally proposed in WAVE~\cite{Andersen:EECS-2017-234}.  
WAVE uses smart contracts and blockchain as a global ledger for all authorizations, 
Delegation of Trust (DoT), and revocations, guaranteeing that all participants 
know the current state of all permissions. 
The same transparency is achieved by our work. 
WAVE also supports out-of-order and non-interactive delegation, 
which is replicated in our smart contracts.
%new wave
In \cite{andersen2019wave}, the authors' experiments
reveal that a Blockchain-based access control system will not scale to a
global size. They state that Blockchain introduces about a minute of latency when adding objects to Ethereum. However, with a private blockchain for a small number of buildings, our solution works well as described in our implementation since managing permissions is infrequent as compared to accessing permission data. Also, the calendar application helps provide access well in advance, reducing the effect of latency. 
To address the scalability issue that may arise in a large campus 
comprised of several commercial buildings, 
other access control methods could be considered in future work.

BOSSWAVE~\cite{andersen2018democratizing} builds on WAVE to provide 
democratized access to the physical resources in buildings. 
Our system is similar to WAVE in that they both leverage blockchain technology. 
However, using smart contracts for authorization to spaces and doorways apart from 
building subsystems and equipment is novel in this work. 
We develop APIs to help interact with the smart contracts;
this will define fine-grained access control for any type of user.

%using APIs and decentralized applications to create the interaction between RDF schema and authorization through smart contracts is novel in our work.

%Our work shows how using decentralized applications (dApp)
%it is possible to make an interactive user interface.
% It also uses the dApp over the decentralized blockchain network 
% to make the system more user-friendly. 
%The interactions of the dApp with the smart contracts and the resource RDF graph 
%could open opportunities for further application development. 
%-----------------------------------------
\section{Methodology}\label{sec:method}
%-----------------------------------------

Our access-control methodology involves four steps:
(1) creating a unified RDF graph of a building by aligning the building's BOT and Brick models;
(2) identifying all possible paths between two locations using a graph traversal algorithm which is implemented by a sequence of SPARQL queries; %DFS
(3) determining the cost of each path by running a number of SPARQL queries; 
(4) granting, revoking, and verifying user permissions to access rooms and equipment therein using smart contracts.
These steps are implemented using three services 
%built on top of a building's timeseries and metadata stores
~\cite{Haggerty13}:
\begin{itemize} %[label=\alph*)]
    \item \textbf{Building-Representation Service} determines the cost of an indoor-path using information captured in the Brick model. % offline, cache the costs of each location
    \item \textbf{Path-Planning Service} relies on the BOT model to find all possible paths between two given locations. These paths are presented to the user along with their costs, enabling the user to choose a desired path. 
    \item \textbf{Access-Control Service} specifies a smart contract, given a delegator, a delegate, a path corresponding to a sequence of building spaces and doors between them, and a time period during which the delegate should be able to access the resources on this path. It also handles validation of a delegate's authority to access a door, at run-time.
\end{itemize}
Below we explain our methodology using an application scenario and discuss how the above services facilitate this application.

% explain the sequence of execution for the services, offline or online

%-----------------------------------------
\subsection{An Illustrative Application Scenario}\label{sec:scenario}
%-----------------------------------------
Consider a simple application scenario, where a group of individuals are invited to a meeting that takes place in a specific room of an office building. The process starts with a calendar invitation created by the meeting host, listing the invitees using their email addresses, the room where the meeting is to be held, and the time when it is to take place. 
The first step is to examine whether the invited individuals are known to the access-control service: if not, new entities need to be created to represent their credentials (Section~\ref{method3}). 
Next, the possible paths in the building to reach the meeting room are computed (Section~\ref{method1}). In this paper, we only consider the problem of indoor path planning, from a main building entrance to a meeting room; nevertheless, our methodology can be extended to the overall problem of planning a path from each individual's location, taking into account their route preferences and parking needs, as long as the relevant information is captured by some geospatial model.
Each of these indoor paths is presented to the meeting host along with a cost that represents the overall risk of giving the invited individuals access to the rooms located on this path (Section~\ref{method2}) and the building elements (equipment, sensors, and setpoints) that contribute to it. The host reviews the possible paths and selects the preferred pathway, which is by default the path that has the lowest cost.
The selected pathway is then transformed into a smart contract (Section~\ref{method3}), where the meeting host delegates each invitee with the necessary privileges to access the various doorways on the path, and a personalized QR code (akin to a boarding pass) is generated to represent each contract. These QR codes are shared with each participant in the calendar invitation.

Before the meeting occurs, each meeting participant arrives to the building and scans their individual QR code to open doors on the path from the building entrance to the designated meeting room. Each QR code scan invokes a request to the Ethereum ledger (Section~\ref{method4}) to verify that the bearer is authorized to access the corresponding door at the present time, given that they have already opened some other doors in a the order pre-specified by the indoor path selected by the meeting host. If the meeting host has access to these rooms and had authorized the meeting participants to access these rooms before the meeting, the access is verified and a control signal is sent to the actuator, i.e., the electric door strike, to open the door. The process repeats for each door along the path. We elaborate on these steps in the following.

%==========================================
\subsection{Building Modeling for Access Control}\label{method1}
%==========================================
The first step of our methodology is to create the BOT and Brick models of the building (if they do not exist already) and align them to create a RDF graph queryable via SPARQL. This RDF graph represents the relationships defined in both Brick and BOT. It allows for identifying pathways in the building using a graph traversal algorithm  and quantifying the cost-sensitivity of each pathway. 

\begin{figure}[t!]
\begin{center}
\includegraphics[width=\textwidth]{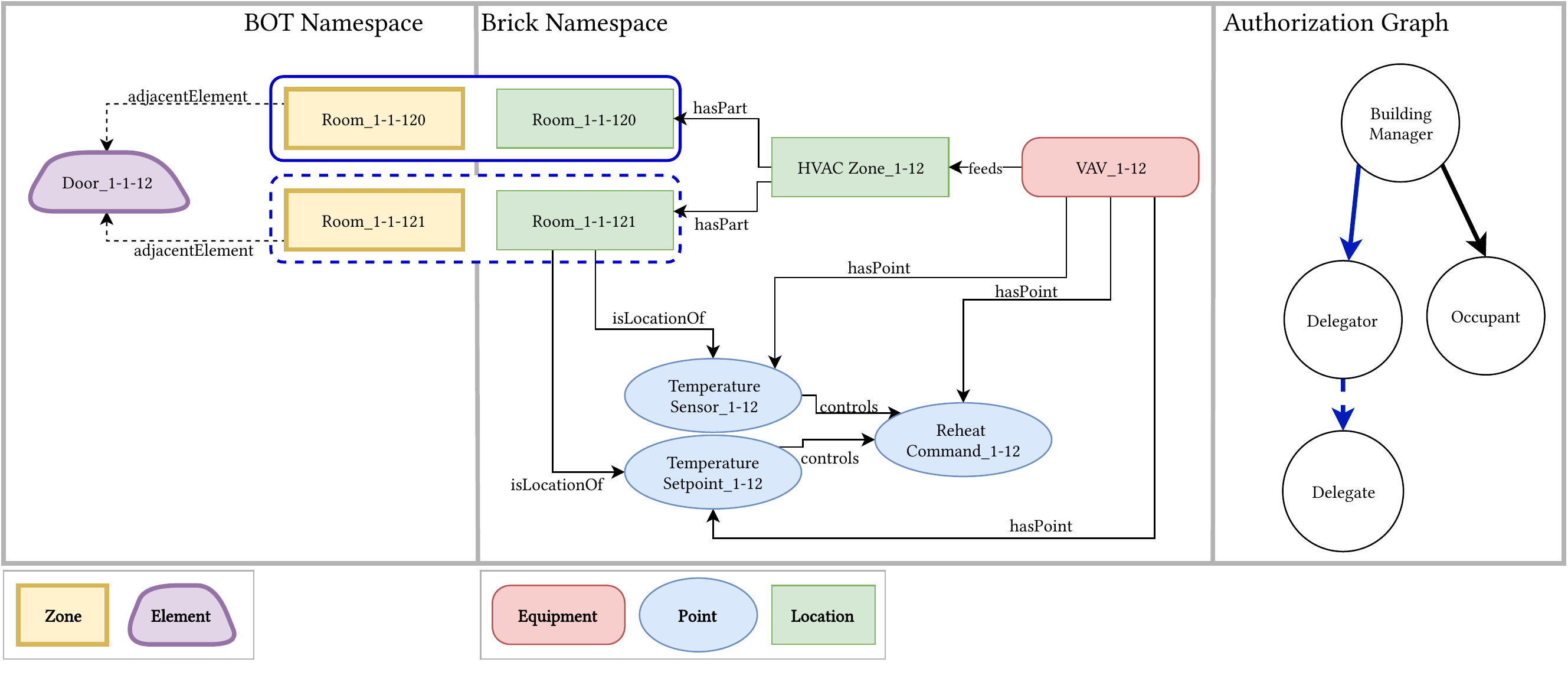}
\caption{The process of aligning Brick and BOT models.
A sub-graph of the the Building Graph Model that captures a complete use case in turtle format can be found in Appendix~\ref{appendix:graph}.}
\label{fig:infer}
\end{center}
\end{figure}

\begin{figure*}[t!]
\begin{center}
\includegraphics[width=.9\textwidth]{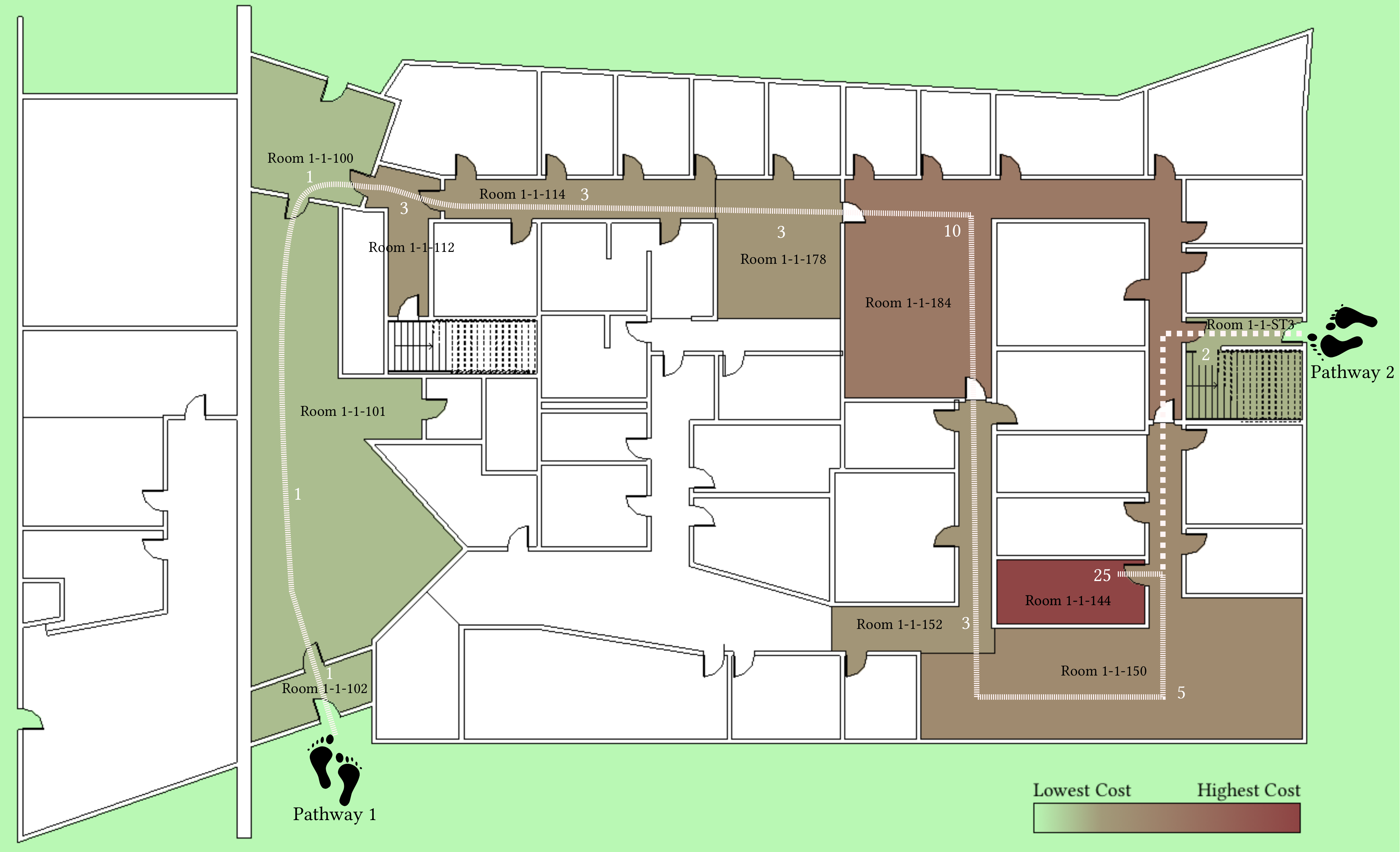}
\caption{A Building's floor plan showing 2 example pathways to a meeting room 
and the cost-sensitivity of every room on these pathways.}
\label{fig:building}
\end{center}
\end{figure*}

Figure~\ref{fig:building} depicts the floor plan of an example commercial building 
occupied by a single organization. We manually converted this floor plan into a RDF graph model of the building based on the BOT ontology\footnote{Generating this model 
can be automated by using information from building information modeling software.}. This model describes the building's topology in terms of where rooms, doors, walls, and other physical elements are located and their adjacency relations. The left panel of Figure~\ref{fig:infer} shows a small subset of nodes in the resulting RDF graph: two rooms of type \textsc{Zone} (subclass of \textsc{Space}) and a door between them of type \textsc{Element}. Each room is connected to the door with the \textsc{adjacentElement} relationship.
%and to the other room via \textsc{adjacentZone} relationship.

Similarly, we created the Brick model of the building in RDF. This involves extracting point names and their types from the BMS, inferring functional relationships between different equipment, and identifying their location from blueprints of the building.
The middle panel of Figure~\ref{fig:infer} shows a small subset of nodes in the resulting RDF graph: two rooms and a HVAC~zone of type \textsc{Location}, a VAV system of type \textsc{Equipment}, and a temperature sensor, a temperature setpoint, and a reheat command of type \textsc{Point}. The VAV system is connected to the HVAC~zone with the \textsc{feeds} relationship, as it supplies air to this zone. The HVAC~zone is comprised of the two rooms, so it is connected to them with \textsc{hasPart} relationship. The temperature sensor and setpoint are located in one of these rooms and are therefore connected to it via \textsc{isLocationOf} relationship. The reheat command is computed based on the difference between the measured and setpoint temperatures and is used to actuate the VAV system; thus, the temperature sensor and setpoint are connected to the command via \textsc{controls} relationship, and the VAV is connected to all three of them via \textsc{hasPoint} relationship.

Once the Brick and BOT models are created, the next step is to align the two models that is to ensure the entities corresponding to the same building location (e.g., room) in Brick and BOT models are the same in both sets of triples. Otherwise, it would be impossible to reason about which pathway enables access to which equipment, which is necessary for establishing our cost function as discussed in Section~\ref{method2}. To this end, we identify syntactic entities that represent the same semantic entity in the two graphs and join them to create a new entity. This new entity is a subclass of \textsc{Location} in Brick and \textsc{Space} in BOT and therefore can be connected to entities defined in both Brick and BOT. The borders around \emph{Room\_1-1-120} and \emph{Room\_1-1-121} in Figure~\ref{fig:infer} show the new entities created by merging respective entities in Brick and BOT models.

%==========================================
\subsection{Quantifying Sensitivity of a Space}\label{method2}
%==========================================

The issue of facility security is quite complex, especially when buildings house expensive or sensitive equipment. Intuitively, the characterization of the sensitivity of a space in terms of standard zones defined in Section~\ref{securityzones} considers the activities taking place in the space and possibly the role of its occupants, but it does not consider the equipment housed in, or accessible through, the space. This is why, our access-control methodology proposes a composite sensitivity function for each space that integrates information about 
(a) equipment and subsystems a user may be able to potentially control by accessing this space,
(b) sensor readings a user may be able to read by accessing this space, and 
(c) the security zone classification of this space according to Table~\ref{table:zones}. 
Through this sensitivity quantification, we aim to help a delegator, such as the meeting host in our example scenario, make informed decisions about providing access to the building spaces.

The RDF model of the building can be used to annotate each building {\em location} with its {\em sensitivity level}. We assume that there are some broadly shared and agreed-upon principles for quantifying the sensitivity of locations. This is a realistic assumption in this domain; in our work, we have adopted the `Hierarchy of Zones' as described in the `Operational Security Standard on Physical Security' of the Government of Canada\footnote{\url{https://www.tpsgc-pwgsc.gc.ca/esc-src/msi-ism/chap4-eng.html}}. We note that similar specifications exist in several other countries, such as the United States\footnote{\url{https://www.esd.whs.mil/Portals/54/Documents/DD/issuances/dodm/522022M.pdf}} and New Zealand\footnote{\url{https://www.protectivesecurity.govt.nz/physical-security/understand-the-physical-security-lifecycle/design/apply-good-practices/security/}}.

%==========================================
\subsubsection{Defining Sensitivity of Security Zones}\label{securityzones}
%==========================================

\begin{table}[t!]
\caption{Zones and their corresponding requirements} 
\begin{adjustbox}{max width=\textwidth}
\begin{tabular}{|l|c|c|c|c|c|}
\hline
\multirow{2}{*}{\textbf{Requirements}}  & \multicolumn{5}{c|}{\textbf{Security Zones}} \\ \cline{2-6}
                      & Public (0) & Reception (1) & Operations (2) & Security (3) & High Security (4) \\ \hline
Monitoring            &        & $\times$         & $\times$          & $\times$        & $\times$             \\ \hline
Screening Required    &        &           & $\times$          & $\times$        & $\times$             \\ \hline
Clearly Separated     &        &           &            & $\times$        & $\times$             \\ \hline
\end{tabular}
\end{adjustbox}
\label{table:zones}
\end{table}

The Canadian standard defines five zones, as seen in Table~\ref{table:zones}. Access to {\em public} zones, such as the grounds surrounding the building, do not need to be controlled. {\em Reception} areas may be inaccessible to visitors, except during specific times of the day or for specific reasons. Access to {\em operations} zones is limited to personnel who work there and to properly escorted visitors. {\em Security} areas are limited to authorized personnel and to authorized and properly escorted visitors. Finally, {\em high-security} areas are limited to authorized, appropriately-screened personnel and authorized and properly-escorted visitors. The standard advises that both {\em security} and {\em high-security} zones should be monitored 24 hours a day, and that zone levels should be accessed in order, i.e., a high-security zone can only be accessed from a security zone. 

%ES: we need an example of a reception and an office 
We use the Zone entity in BOT to model different security zones that exist in a building.
Consider for example the rooms 1-1-100, 1-1-101, 1-1-102, 1-1-112, 1-1-114, 1-1-144, 1-1-150 in Appendix~\ref{appendix:security_zones}. 
Rooms 1-1-102, 1-1-101, 1-1-100 are considered as reception zones, since they are the locations that visitors can access when coming from a public zone without any credentials. 
Room 1-1-112 is considered an operations zone, as only employees are allowed to access the area with proper credentials.
Connected to room 1-1-112 is room 1-1-114, which is considered to be a security zone, as it is physically restricted from room 1-1-112 and additional credentials are required to access the area. 
It should be noted that room 1-1-114 must be accessed from room 1-1-112, as a security zone should only be accessed through an operations zone. 
Room 1-1-144 is an example of a high-security zone that must be accessed from room 1-1-150 which is classified as a security zone.

% Before explaining our composite sensitivity-cost function, 
% let us describe how we quantify each of the above three elements. 
To quantify the sensitivity of the five security zones mentioned above, we map them to an ordinal scale of 0 to 4, with 0 being a public zone, which is not at all sensitive, and 4 being a high-security zone, where access should be carefully controlled. We choose the ordinal scale because the order of security zones signifies their relative importance (e.g., a high-security zone, labelled 4, is more important than a public zone, labelled 3). Thus, each room/space in the building is associated with the cost of its zone. 
%ES: have we discussed an example already with the figure? if not we have to here

%==========================================
\subsubsection{Defining Sensitivity of Equipment and Points}
%==========================================
Building subsystems and points should have different sensitivity costs assigned to them. This is because some subsystems may control more critical aspects of a building (e.g., the lighting system is less critical than the HVAC system), and some points in a subsystem may be more important than others (e.g., using the thermostat to adjust the temperature setpoint is more impactful to the building occupants than simply reading the value of a temperature sensor). 

To account for the fact that some equipment and points are more sensitive than others, we used the Analytic Hierarchy Process (AHP)~\cite{saaty1990analytic} to create a suitable scheme for weighting the sensitivity of each type of equipment and sensing/control points. AHP is a decision-making technique that can be used to prioritize the attributes relevant to a decision-making task: by pairwise comparing these attributes, it helps stakeholders decide on the importance of each attribute relative to others. It has been used for a wide variety of applications, including assessing risk in operating pipelines~\cite{doi:10.1061/(ASCE)1527-6988(2003)4:4(213)} and quantifying the overall quality of software systems~\cite{7748957}. 
In principle, this process should be undertaken by facility-management personnel in collaboration with building owners and occupants. For this work, we answered the questions specified in Appendix~\ref{ahpqa} on the scale defined in Appendix~\ref{ahprating} to develop a list of weights for all types of points (sensors and setpoints) in our model. These weights are shown in Table~\ref{AHPtable}.
%ES: add table here
% assign weights for each type of point and sensor (table)
\begin{figure}
\centering
\begin{minipage}{0.4\textwidth}
\centering
\vspace{3.5em}
\begin{minted}{sparql}
SELECT ?entity ?x
WHERE {
    ?entity rdfs:subClassOf ?x .
    ?x rdfs:subClassOf brick:Point . 
}
\end{minted}
\caption{We obtain all the point types based on the following SPARQL query which obtains the first sub-class of the brick point class in the graph model}
\label{fig:prob1_6_1}
\end{minipage}
\hspace{3.5em}
\begin{minipage}{0.5\textwidth}
\centering
\captionsetup{type=table}
\caption{Weights of the two types of points modeled: sensors and setpoints.}
\begin{tabular}{|l|l|}
\hline
\textbf{Sensors}         & \textbf{Weight} \\ \hline
Temperature\_Sensor      & 0.347           \\ \hline
Damper\_Position\_Sensor & 0.204           \\ \hline
Occupancy\_Sensor        & 0.246           \\ \hline
Humidity\_Sensor         & 0.204           \\ \hline
\textbf{Setpoints}       &                 \\ \hline
Temperature\_Setpoint    & 0.413           \\ \hline
Humidity\_Setpoint       & 0.260           \\ \hline
Air\_Flow\_Setpoint      & 0.328           \\ \hline
\end{tabular}
\label{AHPtable}
\end{minipage}
\end{figure}

In addition to its type, the sensitivity cost of a particular piece of equipment or a sensing/control point depends on what other, and how many, physical components they could impact. This intuition is illustrated in Figure~\ref{fig:infer}. 
In this figure, {\em Temperature\_Setpoint\_1-12} is a {\em Point} element of type {\em Setpoint} and {\em Temperature\_Sensor\_1-12} is a {\em Point} element of type {\em Temperature\_Sensor}. 
These are points of the {\em VAV\_1-12} {\em Equipment} and are both located in {\em Room\_1-1-121}.
{\em VAV\_1-12} feeds fresh air into {\em HVAC Zone\_1-12} which is of type {\em Location}.
{\em Room\_1-1-120} and {\em Room\_1-1-121}, which are also of type {\em Location}, 
are parts of the {\em HVAC Zone\_1-12}. 
In estimating the sensitivity of this room, 
we argue that one would have to take into account the fact that 
VAV equipment can be impacted by the actions of the room occupants, 
who may read the {\em Temperature\_Sensor\_1-12} value and control the {\em Temperature\_Setpoint\_1-12}.
Thus, {\em Room\_1-1-121} is more sensitive than {\em Room\_1-1-120} 
and occupants need to be aware of this when granting permissions to other occupants and visitors.

% provide a formalized notation of how access is quantified
% follow this paper: https://www.researchgate.net/profile/Adriana_Cherri/publication/318767781_Route_optimization_in_mechanized_sugarcane_harvesting/links/5b3154c44585150d23d44416/Route-optimization-in-mechanized-sugarcane-harvesting.pdf
% something about pathways and how this we are trying to optimize for lowest costs

%==========================================
\subsubsection{Defining Sensitivity of an Indoor Path} \label{sec:queries}
%==========================================
Finally, the overall sensitivity cost of a potential indoor $path$ is calculated as the sum of all the costs of the rooms $r$ it includes based on their security zone classification, plus the costs of all points $p$ located in the rooms $r$. 
The total cost of each point $p$ is initially assumed to be the value of 1, and increases with the number of rooms $r$ affected by point $p$. Furthermore, the cost is scaled by $\text{weight}(p)$, the sensitivity of the point as established through the AHP process and seen in Table~\ref{AHPtable}. This intuition is captured by the following function:
\begin{align}
\text{cost}(path) &= \sum_{r \in path} \text{cost}(r)\\
                  &= \sum_{r \in path} \left(\text{sensitivity}(r) + \sum_{p~\text{hasLocation}~r} \text{weight}(p)\times(1 + control(p))   \right),
\end{align}
where $path$ is a sequence of rooms $r$, 
$\text{sensitivity}(r)$ is the numerical value of security zone classification for room $r$,
$p$ is a point (e.g., a setpoint or a sensor) that is part of a subsystem,
$\text{weight}(p)$ is the weight given to point $p$ (determined by AHP),
and $\text{control}(p)$ is the number of locations or zones affected by point $p$.
For example, in Figure~\ref{fig:infer}, we see that {\em Temperature\_Setpoint\_1-12} has $\text{control}(p)=2$ as ambient air temperature of two rooms would be affected by adjusting this setpoint. A complete calculation for the costs of \emph{Pathway 1} and \emph{Pathway 2} can be found in Appendix~\ref{pathways}.
%whereas \em{Room\_1-1-120} has $\text{control}(p)=0$ since the points of the location does not control any points.

% \begin{align}
%         control(p) &= \left\{\begin{matrix} n, & if~n~\text{locations that}~p~\text{affects with control} \\ 0, & \text{otherwise} \end{matrix}\right.
% \end{align}

It is important to explain here the role of the building model in computing the path-sensitivity cost function above. We have developed a set of SPARQL queries based on the building's RDF model (see sub-graph in Appendix~\ref{appendix:graph}) to compute: 
\begin{itemize}
    \item the sequence of adjacent {\em rooms} leading from one location to another, e.g., the main building door to the meeting room; and
    \item the set of {\em points} located in a room, their types, and the locations they influence through control.
\end{itemize}

\begin{flushleft}
We seek to answer the following questions with SPARQL queries in the namespaces found in Appendix~\ref{appendix:namespace}:
\end{flushleft}
\medskip
\begin{flushleft} \textbf{Q1 -- what are the possible sequences of adjacent resources (locations or doors), starting at {\em main entrance} and reaching a specific {\em meeting room}?}
\end{flushleft}
\smallskip
\begin{minted}[linenos]{sparql}
SELECT ?element 
WHERE { ?element bot:adjacentElement <Resource>. }

SELECT ?element # reverse query of lines 1-2
WHERE { <Resource> bot:adjacentElement ?element. }
\end{minted}
\smallskip
\begin{flushleft}
The two queries above are repeatedly invoked, starting with the building's {\em main entrance} as $<Resource>$ to find an adjacent door or room $?element$ using the query from lines 1-2. For each of the $?element$s, if $?element$ is a door, then the query from lines 3-4 is executed next using $?element$ as $<Resource>$ to find more doors or rooms. Otherwise, the query from lines 1-2 are repeated with the $?element$ as the new $<Resource>$. This process is repeated until the destination {\em meeting room} $<Resource>$ has been reached. 
Note that when a query returns multiple rooms or doors, we need to repeat this process for each of them separately.
Using these two queries allows the Path-Planning Service to perform a recursive depth-first search by finding adjacent resources to each `Room'.
\end{flushleft}
\medskip
\begin{flushleft}
\textbf{Q2 -- what is the sensitivity of a {\em location} based on its {\em security zone} classification?}
\end{flushleft}
\smallskip
\begin{minted}[linenos]{sparql}
SELECT ?seczone
WHERE { 
    ?seczone bot:hasSpace <location> .
}
\end{minted}
\smallskip
\begin{flushleft}
This query helps to determine the $sensitivity$ of <location> where $?seczone$ is subsequently translated into a number based on its ordinal property described in Section~\ref{method2}. 
\end{flushleft}
\medskip
\begin{flushleft}
\textbf{Q3 -- what {\em points} are in a {\em location}? What kinds of {\em points} are they? Can the points influence other locations through {\em controls} relationship?}
\end{flushleft}
\smallskip
\begin{minted}[linenos]{sparql}
SELECT ?point ?subsubtype ?subtype
WHERE { 
    <location> bf:isLocationOf ?point .
    ?point rdf:type ?type .
    ?type rdfs:subClassOf* ?subsubtype .
    ?subsubtype rdfs:subClassOf ?subtype .
    ?subtype rdfs:subClassOf brick:Point .
}

SELECT ?point ?location
WHERE { 
    <location> bf:isLocationOf ?point .
    ?point bf:controls ?command .
    ?equipment bf:hasPoint ?command .
    ?equipment bf:feeds ?zone .
    ?zone bf:hasPart ?location .
}
\end{minted}
\smallskip
\begin{flushleft}
The first query from lines 1-7 determines the points that are located in <location> and their types  (subsubtype = Temperature\_Setpoint and subtype = Setpoint). The second query from lines 9-16 decides whether the point influences any other locations by determining the equipment of the command that the point controls and seeing if the equipment feeds any locations. 
These queries are executed by the Building-Representation Service to determine the cost of an indoor-path.
\end{flushleft}

\subsection{Managing Smart Contracts}\label{method3}
%==========================================

Upon selecting an indoor path that meeting participants can follow to reach the meeting room, a smart contract must be specified and run by the Access-Control Service to authorize the participants to open the required sequence of doors and access certain spaces shortly before the meeting starts.
%One of the smart contract was used to create new users and access rules for individuals in the building. The smart contract is also used to revoke access. %ES: weird sentence ....
Users who already have access to (some of) the building's systems and spaces can delegate access to other individuals. This creates a tree-like structure of an authorization graph, as depicted in Figure~\ref{fig:infer}. This implies that the process has to be bootstrapped with some original space manager(s), e.g., the Building Manager in Figure~\ref{fig:infer}. Traditionally, they might be the building-security personnel, or (some of) the long-term building occupants.

% Let us revise our application scenario. % OA: don't need this sentence
%If the host (the delegator) wishes to invite someone (the delegate), then 

In our application scenario, the meeting host (the delegator in Figure~\ref{fig:infer}) is assumed to have access to the whole floor (including \emph{Room\_1-1-120} and \emph{Room\_1-1-121}) and can thus delegate access to the spaces and doorways in the path leading to the meeting room (i.e., \emph{Room\_1-1-120}). Hence, the delegator grants the meeting participant (the delegate in Figure~\ref{fig:infer}) access to all the doors and rooms in the path they have selected, which implies that all the rules mentioned in the smart contracts should be satisfied for the delegation to succeed. 
To that end, the delegator first uses the smart contract to create a new unique entity for the delegate, assuming that she is not already known to the system (otherwise the existing entity is used to add a new access rule). Each entity is a node in the authorization graph, therefore creating a new node for Visitor B.
The delegator provides the delegate's address, which is the public key of the delegate, along with the sequence of permissions implicit in the chosen path and the period during which these permissions should be valid, in effect, a period covering the meeting duration. If no failure is encountered in the creation of entity and all the requirements of the contract are fulfilled, a valid entity is created for the delegate.

Next, a new access rule has to be created, containing the delegator's entity (the meeting host) as the source, the delegate's entity as the target, the expiry time after which the access rule will become invalid, and the list of resources that the delegate is allowed to access. This creates the relationship between two nodes, Occupant A and Visitor B in Figure~\ref{fig:infer}, i.e., the delegator and the delegate, 
which describes the access rule.
The list of resources received from the Path-Planning Service are stored in the order they should be invoked, starting from the building's main entrance and concluding with the meeting room. At run time, the in-order access of the list of locations and doors is evaluated, which helps the delegate to navigate in the right path. %, and to not attempt any unauthorized access. 
The destination and equipment that can be accessed by the delegate are also stored.

Once all the requirements to create a new access rule in the smart contract are fulfilled, a successful transaction is executed. A failure would revert any changes that were made to the contract's data.

An important advantage of smart contracts is the flexibility they afford in the revocation process. Once a smart contract has been issued to authorize a delegate with access to some spaces, it cannot be deleted; the blockchain is immutable and anything stored on the blockchain as a transaction cannot be deleted. To revoke this authorization, the access rule assigned to that user must be made invalid. Similarly, to remove a user from accessing any system or space in the building, for instance when an employee leaves the organization, the entity belonging to that user is made invalid. This is enough to make sure that this user can no longer access building spaces and equipment therein.
Once a rule or entity is made invalid, it can no longer be reused for the same address of the delegate.

%==========================================
\subsection{Checking Credentials at Run-Time}\label{method4}
%==========================================
Upon arrival, when the delegate tries to access any space or equipment, the data from smart contracts need to be read by the Access-Control Service to validate whether the delegate's credentials authorize them to access the spaces they attempt to access. 
To that end, the access rule assigned to the delegate is read from the blockchain. From this data, the duration and validity is first checked. If the access rule is valid, the set of accessible resources assigned for the delegate are checked and the permissions granted to the delegate, as a part of the access rule, are verified. If the requested action is included in the set of permissions and the requested resource is included in the set of accessible resources, then the user will be allowed to access the requested resource. 
Similar is the case for accessing points and equipment. In order to read or modify the temperature setting in the room, the accessibility needs to be checked and the smart contract decides if the requested action can be taken.

%As in the above example, the delegate's access rule has only room number 112 added to the set of accessible resources. If the visitor tries to perform the action of entering the space, room number 112, the contract will verify and let this visitor access the room. If she tries to enter room number 113, the contract will return negative accessibility value and she will not be able to enter this space.
%-----------------------------------------
\section{Implementation}\label{sec:implement}
%-----------------------------------------

In this Section, we describe the prototype implementation of our methodology on our own private blockchain using the Ethereum network.
%, since all the elements of the prototype are more dimensioned providing us more accurate results than a public blockchain when evaluating the system. However, ultimately, this implementation has to be deployed in public blockchains in real scenarios.
Ethereum is a blockchain platform that includes a Turing complete scripting language called {\em Solidity} for building, deploying, and implementing smart contracts. These contracts have no restrictions in terms of size and are stored in the blockchain. 
Our methodology relies on three smart contracts, namely {\em Archives}, {\em Implications} and {\em Exclusions}. They are defined as follows:
\begin{itemize}
    \item The \textbf{Archives} contract manages and stores the entities and access rules for each user. Any creation or deletion of entities or access rules leads to a state change. The state of the Archives contract is read to verify access.
    \item The \textbf{Implications} contract cross-verifies the validity of access rules and processes the path resources that should be accessible to an entity for a specific access rule. A state change takes place only when a new rule is created.
    \item The \textbf{Exclusions} contract is used when the delegator chooses to provide a list of all the resources that the delegate is not allowed to access, even though they are in spaces that the delegate is allowed to reach. 
\end{itemize}

These smart contracts are compiled and deployed using scripts. For each contract, a unique bytecode and contract address is created. The contract address is used to access a contract in the private blockchain network, either from external functions or from another smart contract. An API, illustrated in Figure~\ref{fig:three-apis}, enable the client applications, such as our meeting-planning example scenario, to interact with the deployed smart contracts. The API exposes a number of operations, receiving (or producing) JSON objects as input (or output). 

% \begin{figure}[!h]
% \begin{center}
% \includegraphics[width=.7\textwidth]{architecture1.pdf}
% \caption{Overall System Architecture}
% \label{fig:sc}
% \end{center}
% \end{figure}

%==========================================
\subsection{Adding a New User}
\begin{figure}[!ht]
%ES change the figures to be delegator and delegate; number each horizontal line (call) and refer to their numbers in the text when you explain the workflow
\centering
\subfloat[][Adding a new Entity.]{\includegraphics[width=.95\textwidth, height = 3.5cm]{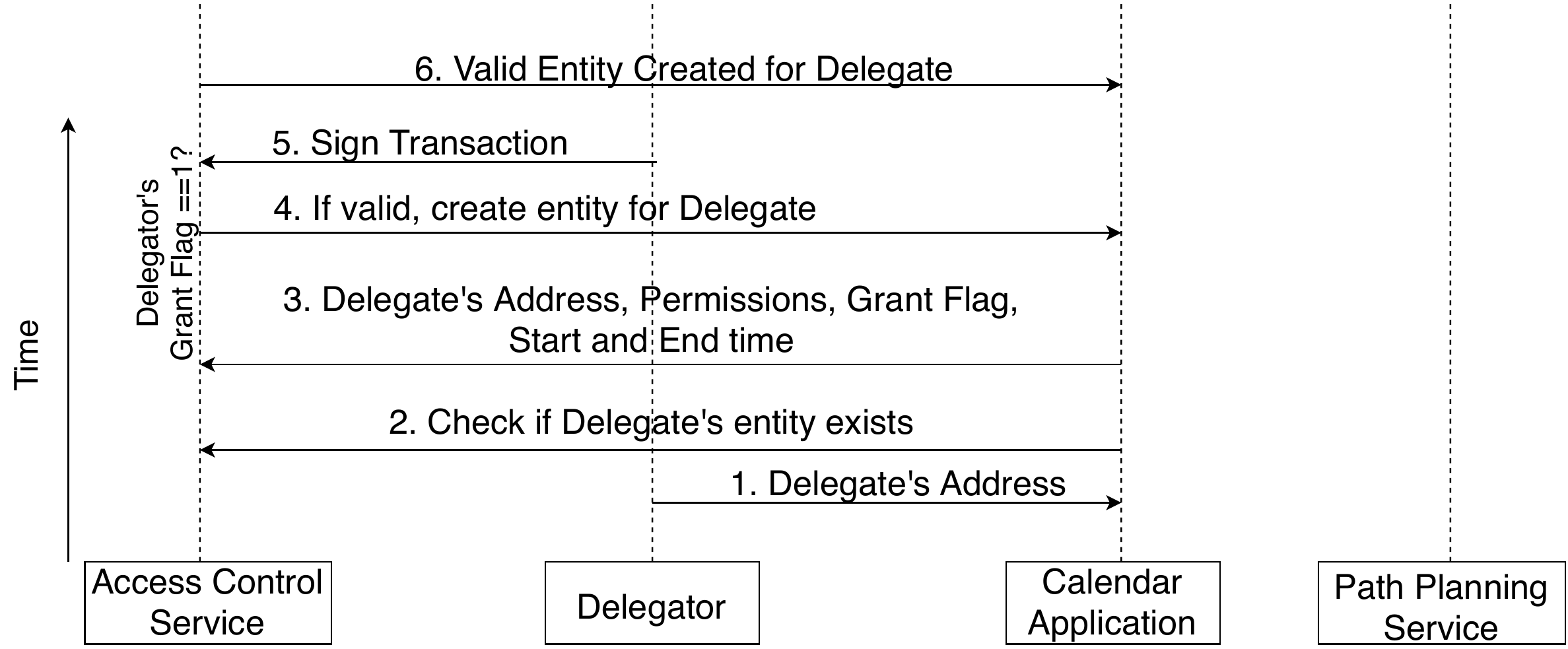}\label{fig:newEnt}}\\
\subfloat[][Adding a new Access Rule.]{\includegraphics[width=.95\textwidth, height = 3.5cm]{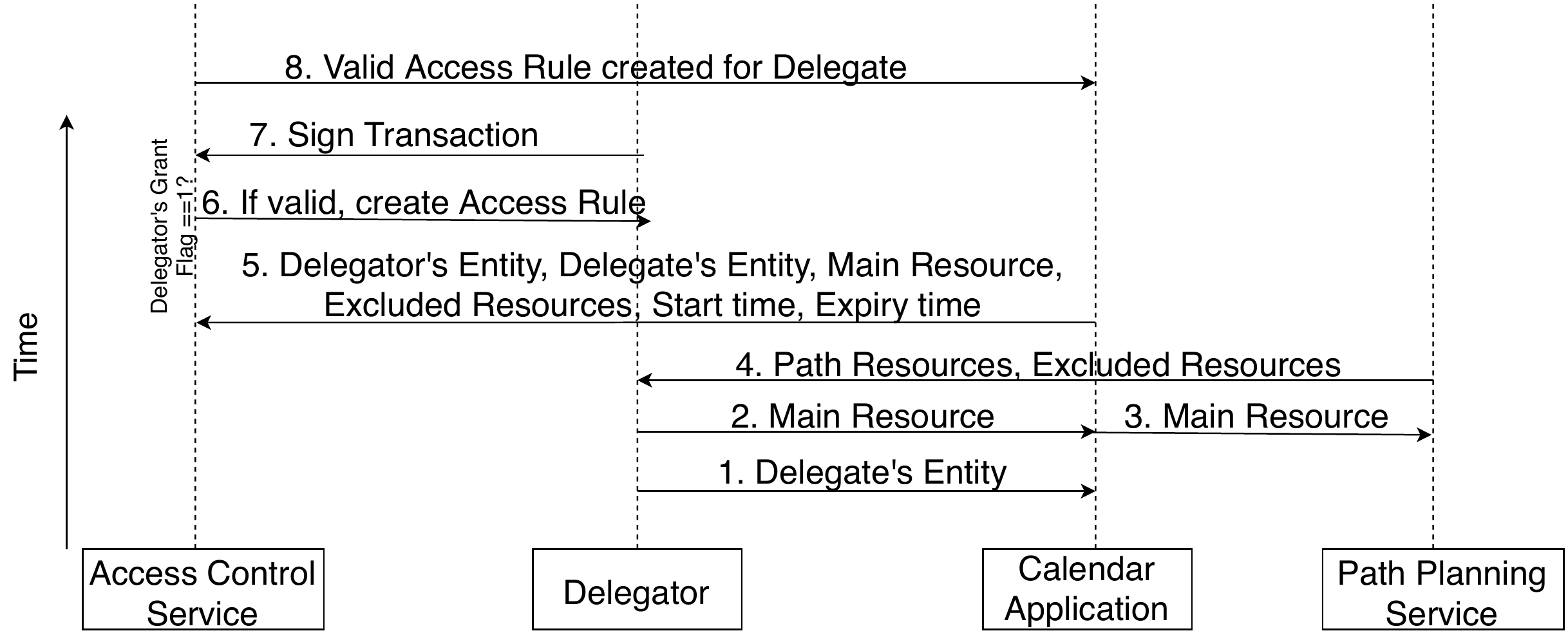}\label{fig:newRule}}\\
\subfloat[][Verifying an Access Rule for accessibility information.]{\includegraphics[width=.95\textwidth, height = 3.5cm]{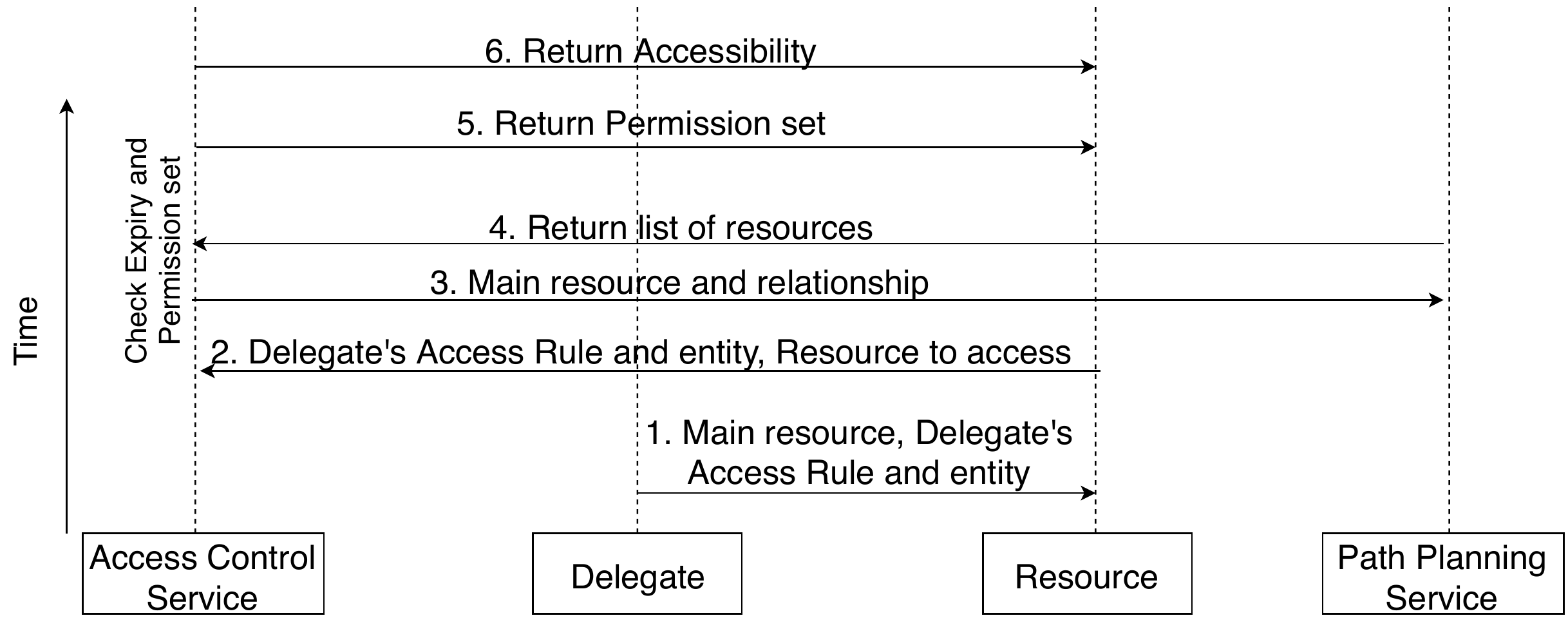}\label{fig:check}}
\caption{Three APIs for smart-contracts}\label{fig:three-apis}
\end{figure}

%==========================================
As shown in Figure~\ref{fig:newEnt}, the delegator provides the address of the delegate to the meeting-planning application to create a calendar invitation. This address is used to verify if an entity exists already for the delegate, or a new entity needs to be created by the Access-Control Service\footnote{
Note that there are commercial solutions for decentralized identity management 
such as Civic (\url{https://www.civic.com/}), Sovrin (\url{https://sovrin.org/}), and 
uPort (\url{https://www.uport.me/}). 
These solutions allow for users to manage their own identities, and can be integrated with our system
to eliminate the need for independently managing and verifying the identity of building occupants and visitors.
However, due to the commercial nature of these solutions, we do not use them in our proof-of-concept implementation. 
As a future direction, it would be interesting to see the performance impacts of using 
a third-party decentralized identity management service.}. The start and expiry times for the entity should also be provided by the delegator. If no expiry time is specified, the implication is that the contract is valid for the foreseeable future until it is explicitly revoked. In our application scenario, meeting participants should only have access to the building for a specific time period. Hence, the expiry time must be provided.

In addition to the delegate's address, and start and expiry times, the set of permissions that the delegator intends the delegate to possess must be provided as input to the API. These include read (for accessing sensor measurements) and write (for changing the value of a setpoint) permissions for points and equipment resources. The permissions set also includes a flag to indicate whether the new delegate entity is allowed to further delegate access to other users. 
%Considering that the current delegator has grant permissions, she can create a new entity and access rule for other users. 
%ES does the above say anything new?
%LB its the vertical written text in the figures to check grant flag
% In our scenario, the delegate is a visitor and hence, may not be given permission to grant access to other users.
The meeting-planning application interacts with the Access-Control Service, providing the required information as depicted in Step~3 of Figure~\ref{fig:newEnt}. This information is sent to the Archives contract to create a new entity. The contract checks validity of the input and creates an entity for the given delegate's address.

% The smart contract checks the validity of the input and other rules that are enforced for an entity, 
% like the delegator having a valid grant access. 
Once all the checks of the smart contract are successful,
%ES: what does the above mean?
%LB checks are the rules that the smart contract checks for the validity and permissions-- included in the paragraph above
the delegator has to sign the transaction using their private key. Any failure in the transaction would revert all the changes made. If a valid response is not returned from the smart contract, the entity is not created for the delegate. The API reads the transaction hash received from the contract's callback function and finds out whether a valid entity was created for the delegate. Every entity created has a unique address assigned to it.

%==========================================
\subsection{Adding a New Access Rule}
%==========================================
After creating a new entity for the delegate, the delegator can proceed to create an access rule, based on the spaces and equipment resources that the delegate is authorized to access. 
In our application scenario, the meeting participants must be allowed to access all spaces they need to go through to reach the meeting room, and all sensing and control points located in this meeting room. This is exactly the information that the path-planning queries Q1 and Q3 deliver (refer to the discussion in Section~\ref{sec:queries}).
%in Figure~\ref{fig:sparql} delivers.
%To achieve this objective, the delegator interacts with the Resource Retrieval Service (RRS),providing a starting and ending resource to the service. This service returns all the possible paths that the delegate can take from the main entrance to another space, say Room-1-1-113, along with the sensitivity cost of the paths and lists the spatial resources that should be given access. 
If the delegate is already known to the system and has previously had access to some resources in the building, access must only be provided for the additional spaces that are on the selected path.

% Recall that the Building Representation Service estimates the sensitivity cost of a path,
% considering all sensing and control points located in the spaces that are in the path.
%resource, that is Room-1-1-113. These consists of the resources in the Brick RDF graph that are related to Room-1-1-113 with "feeds" and "hasPoint" relationships. 
The delegator has two options with respect to the various points located in the destination room: (a) they may authorize the delegate to access all of them, except an explicitly excluded set, or (b) they may include all of them in the contract (default behaviour). The Exclusions contract is used to implement the former.
%%%why contracts are not just returning the state
The permissions set mentioned in the previous section determines if the delegate can read sensor measurements or write control setpoints in the destination space. 

The smart contract's function requires information such as the delegator's and delegate's addresses. The contract verifies the entities through these addresses. In addition, the contract checks if the delegator's entity has the permission to grant access to new users.

For each access rule, the Path-Planning Service provides a list of resources for the delegate to access as shown in Figure~\ref{fig:newRule}. This list is generated when the delegator selects a path (likely the one that has the lowest sensitivity cost). The associated list of path resources and inaccessible equipment (that could be specified by the delegator) is stored with the Implications and Exclusions contracts. These contracts identify a unique access rule using its hash. The two contracts are called from the Archives contract when a new access rule is being created to add the path resources and excluded equipment list. 
In our application scenario, the Implications contract defines the order in which the visitor should access the doors to reach the meeting room. But if access is being granted for an occupant, there needs to be no restriction on the order of the path that should be followed. This processing is also done by the contract apart from storing the list.
%%%why contracts are not just returning the state

Other parameters required by the contract's function include the start and expiry times for the access rule being created and the main or destination resource, as in Step 5 of Figure~\ref{fig:newRule}. All the fields are packaged into the request in JSON format. The API runs a callback function to the smart contract to create the new access rule with the provided fields. The access rule is uniquely identified with a hash created using the delegator and delegate's information.
The Archives contract is called to create the access rule. It verifies the entities, checks the granting rights of the delegator, and also validates that such an access rule was not defined earlier and is a new access rule for the delegate.

A valid access rule is created with the required fields as shown in Steps~6-7 
%ES what call is that in the figure?
the delegator signs the transaction using their private key. If the response returned from the smart contract is not a %valid,
success value
%es how is validity checked? --- by the API, mentioned below
then the new access rule for the delegate is not created. The API reads the transaction hash received from the callback function of the contract and responds if a valid access rule was created for the delegate.

%==========================================
\subsection{Verifying User's Access Privileges}
%==========================================
While accessing a resource, the existing entities and access rules are read from the smart contract. It is important to note that the run-time verification of the user's credentials simply queries the private blockchain and does not cause a state change. Thus, it does not require any ``gas'' (i.e., has zero transaction fee). However, creating a new entity for the delegate or adding new access rules costs a specific amount of ``gas'' since they result in state changes on the private blockchain.

When the meeting participant arrives to the building and tries to access a specific resource, e.g., the meeting room, a request is sent to the API endpoint to verify if they indeed are authorized to access this resource. As in Steps 1-2 of Figure~\ref{fig:check}, the API request requires
%ES make sure that it is the same as the diagram - ref the figure here; the sentence below if weird
the entity and access rule of the delegate, the resource name, and the action to perform. The state of the smart contract is read to retrieve the details of the entity and access rule for the delegate. The functions of the smart contract check the validity of the entity and access rule, and verify its start and expiry times. Next, the permission set is inferred from the entity, the main resource, the implication and excluded resources are found from the access rule. This process is described in the Figure~\ref{fig:check}.

After receiving this data from the contract, the requested resource is looked up in the implications and excluded resources. If it is not found, a query is invoked (Q3 in Section~\ref{sec:queries}) to identify all the resources related to the main resource with the required type of relationship, shown in Steps 3-4 of Figure~\ref{fig:check}. If the requested resource is included in the returned resources, then the delegate has access to the requested resource. Otherwise, the delegate is not authorized to access the requested resource.
%ES weird senetnce above and actually inconsistent with the text above... we need towork line by line here

If the resource is equipment or a point in Brick, then the permissions from the delegate's entity are verified to see if they can read or write (control) the requested resource. Once the validation process is completed, the API returns a JSON response indicating whether the requested action is allowed or not. Since these actions do not need to be mined, they are performed immediately without any delay.

Similarly, whenever a visitor enters the building through the main door, they have to follow the path to which the access was provided. 
% As in Figure~\ref{fig:building}, the spaces along the path with the lowest cost
% are typically chosen by the delegator. 
Hence, the delegate has to follow a sequence of doors in order to reach the destination room. For this, the smart contract function reads the set of implications that lead to the destination room. The implications are stored as in the order of the path. This leads to verifying a sequence of access requests.

A user pointer is used to store the current location of the delegate. When a delegate tries to access a resource present in the set of implications, the pointer's current value and the requested resource are verified in the sequence of the path. If the order of requested access matches the path, the delegate is allowed to unlock the door through hardware control of the smart lock. If there is a mismatch, it would imply that the delegate has either tried to enter a wrong door or has skipped a door on the way. This method enforces that the delegate follows the path he is supposed to and not get lost on his way to the destination.

The time to access the implications should also be considered when specifying the start time. 
For instance, access could be provided to the delegate 30 minutes in advance 
so that they can come in early and reach the designated room. 
Requests invoked to check access for a user at different stages are logged. 
In future work, we plan to add new smart contracts for tracking the number of times 
the state of the contract is read to verify access 
to a certain zone, and the users who requested access.
%-----------------------------------------
\section{Evaluation}\label{sec:perform}
%-----------------------------------------

\begin{figure}[t!]\centering
\subfloat[Busy Day Schedule]{\label{a}\includegraphics[width=.32\linewidth]{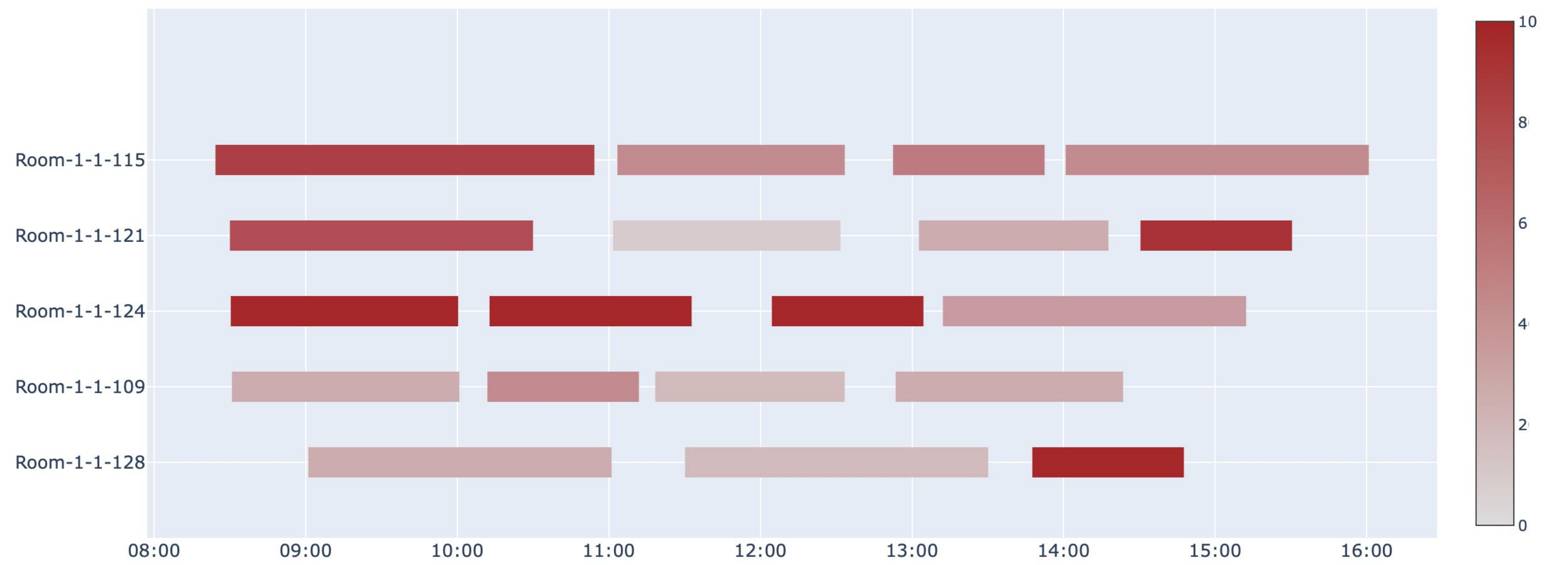}}\hfill
\subfloat[Average Day Schedule]{\label{b}\includegraphics[width=.32\linewidth]{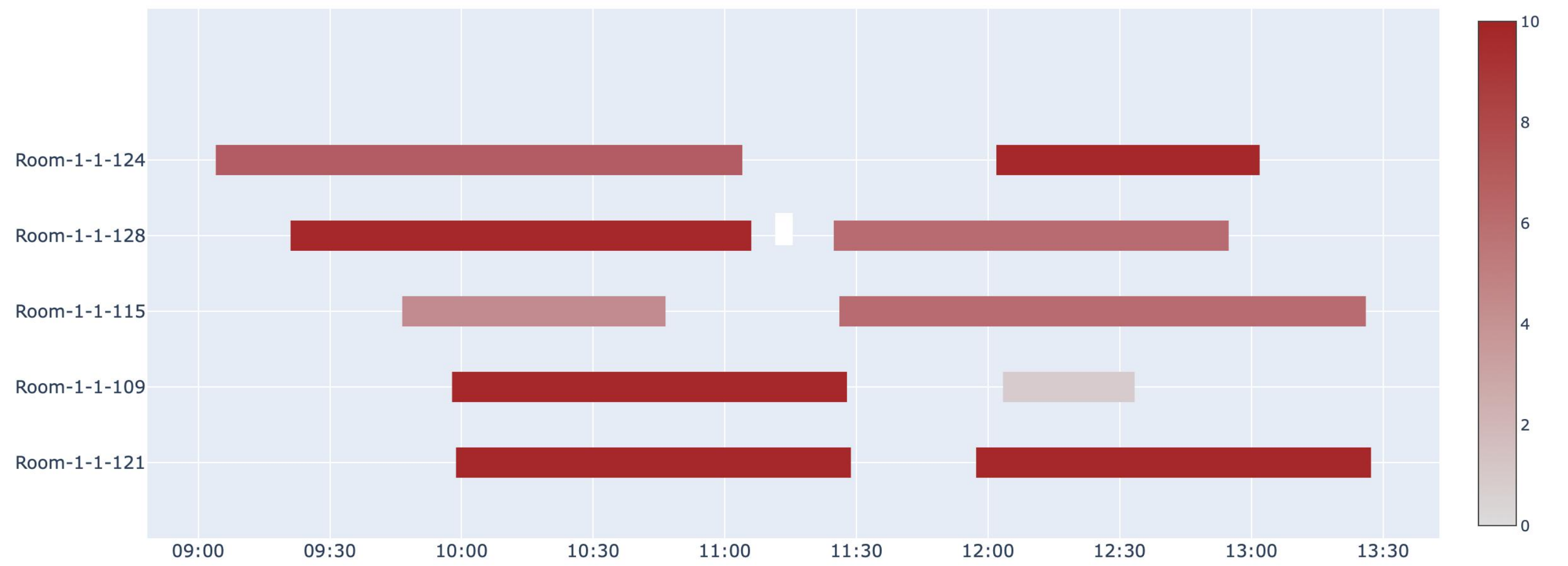}}\hfill
\subfloat[Quiet Day Schedule]{\label{c}\includegraphics[width=.32\linewidth]{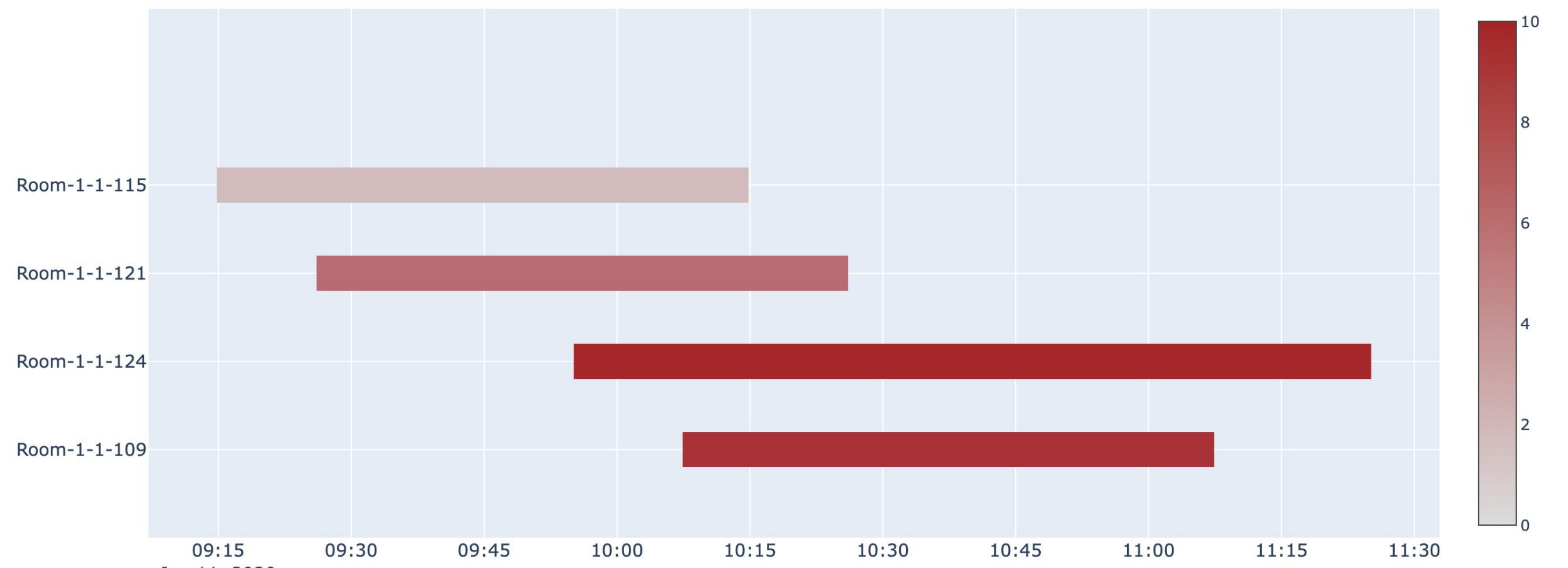}}
\caption{Meeting schedules in 3 different types of days}
\label{fig1}
\end{figure}

\begin{figure}[t!]\centering
\subfloat[Busy Day Occupancy]{\label{a}\includegraphics[width=.32\linewidth]{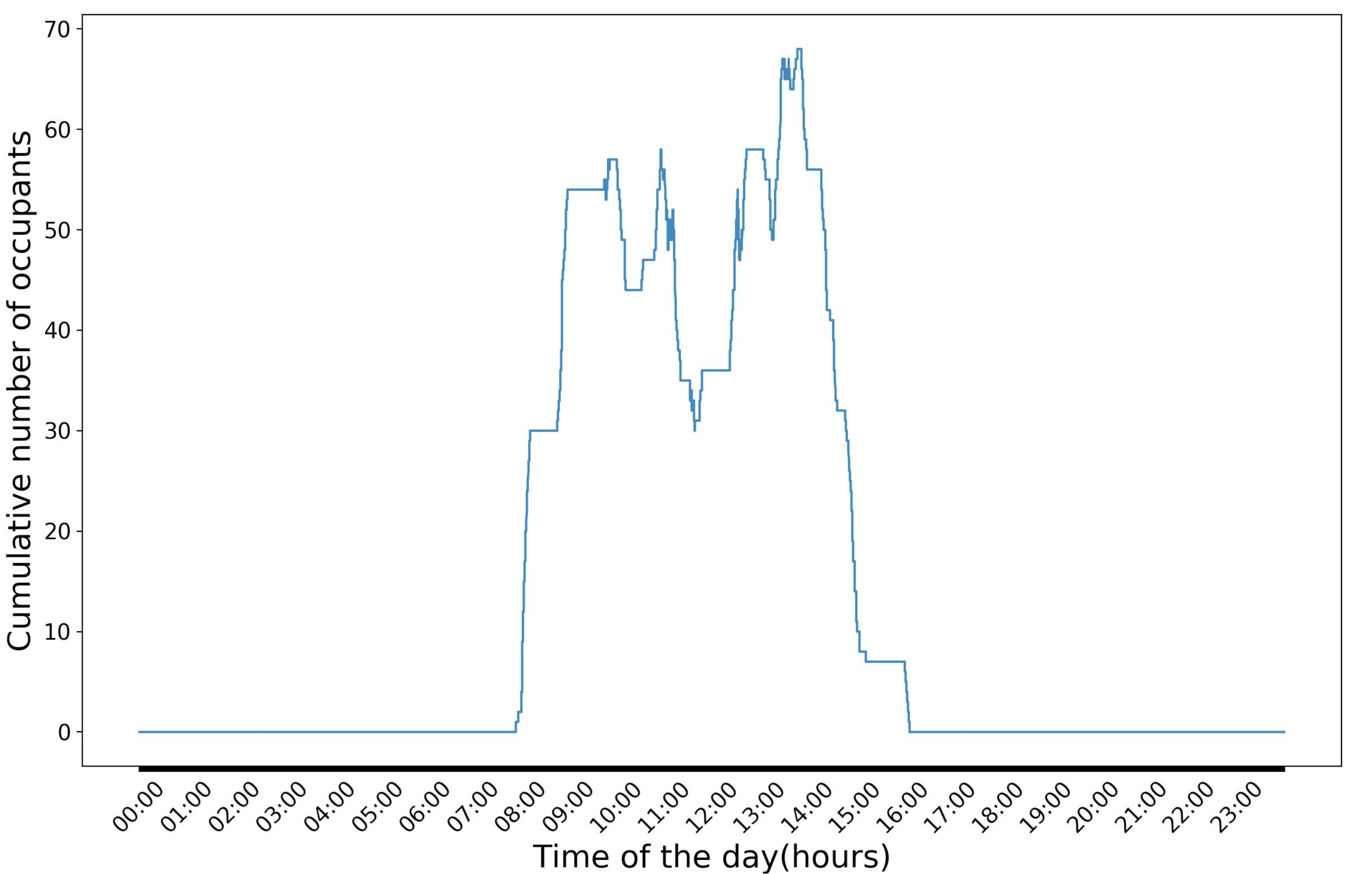}}\hfill
\subfloat[Average Day Occupancy]{\label{b}\includegraphics[width=.32\linewidth]{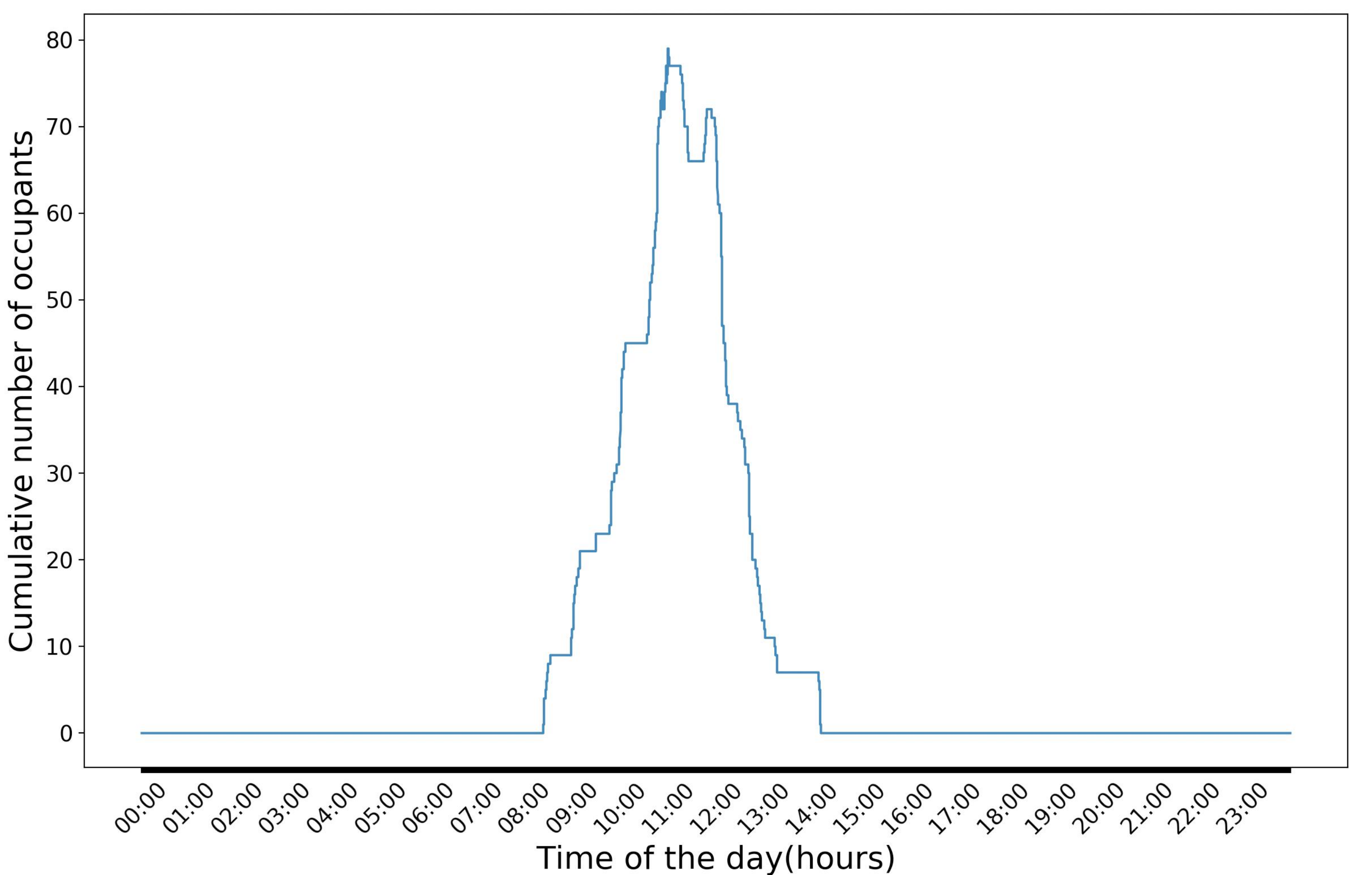}}\hfill 
\subfloat[Quiet Day Occupancy]{\label{c}\includegraphics[width=.32\linewidth]{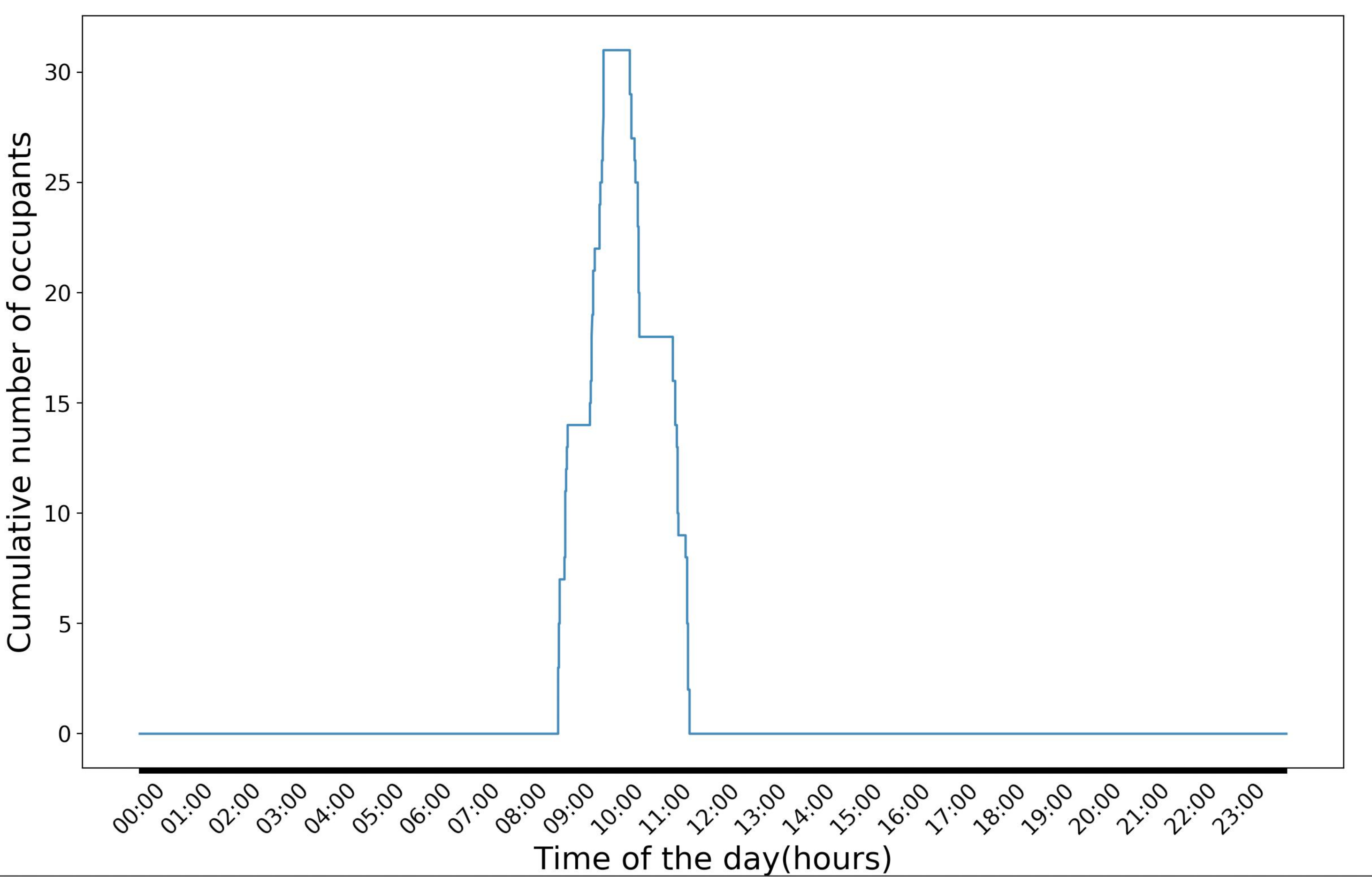}}
\caption{Occupants in the building during 3 different schedules}
\label{fig2}
\end{figure}

We simulate a number of concurrent meetings in the example building depicted in Figure~\ref{fig:building}
to evaluate the performance of the proposed system in realistic scenarios.
This building has five conference rooms that could be used to hold meetings in parallel.
To create different load levels, we consider 3 types of days: busy, average, and quiet days.
We assume that there are more concurrent meetings with more participants on a busy day than an average day or a quite day.
For each type of day, we model the number of meetings held in each conference room by a Poisson process.
Hence, the intervals between successive meetings held in the same conference room are exponentially distributed.
We assume that the duration of a meeting could span from 60~minutes to 150~minutes,
so we sample the length of each meeting from this uniform distribution.

The next step is to model arrivals and departures of meeting participants. 
We consider a window of 30 minutes before the meeting starts and after it ends,
and use a Poisson process to simulate arrivals and departures in these two windows.
Figure~\ref{fig1} shows example schedules for all meeting rooms in the 3 types of days we considered.
The color intensity depicts the number of participants of each meeting; 
the darker the color is the more people attend the meeting.
Figure~\ref{fig2} shows the cumulative number of people in the building throughout the day for the 3 types of days.
We have a maximum of around 70 meeting participants on busy and average days,
and a maximum of 30 meeting participants on quite days.
% Table~\ref{tab:Rtable} summarizes the average length of meetings, 
% the average number of meetings per day, 
% and the average number of participants in a meeting for these schedules. 

When access is delegated to an attendee, the paths to these meeting rooms are received from the path planning service.
Accordingly, we calculate the number of requests a participant makes to receive and verify access. 
The average number of requests is between 4.5 and 5.5 requests per participant in our simulations.

% \begin{table}[h!]
% \begin{tabular}{|D|Q|Q|Q|Q|}
% \hline
% \textbf{Type of day}
% & \textbf{Average Length of meeting}         
% & \textbf{Average number of meetings} 
% & \textbf{Average number of participants} 
% & \textbf{Average number of requests per participant} 
% \\ \hline
% Busy Day      & 86.3 minutes & 14 & 6 & 4.5  \\ \hline
% Average Day& 90 minutes & 14&10&5.5 \\ \hline
% Quiet Day &105 minutes & 4 & 8 & 4.8 \\ \hline
% \end{tabular}
% \caption{Summary of meeting attributes and requests.}
% \label{tab:Rtable}
% \end{table}

%-----------------------------------------
\subsection{Analyzing Delay and Cost}
%-----------------------------------------
Performance evaluation is done on a private blockchain network consisting of 2 mining nodes. 
The API sends simultaneous and parallel requests to these nodes and transactions are executed by the miner nodes. 
%An average block size of 10kB is used to achieve acceptable latency.
We measure the latency of transactions as perceived by the user, i.e., the time between the user issuing a request 
to the access control service and receiving a response from the service. 
As described in Section~\ref{sec:implement}, the access control service interacts with the blockchain mining nodes to issue transactions. 
Table~\ref{tab:Rtable2} describes the average delay for adding a new entity for a user unknown to the system
and adding access rules for each participant of the meetings to the blockchain. 
It also shows the cost associated with transactions to add new entities and access rules. 
At the time of evaluation, one ether was equal to approximately 143.68 USD.

\begin{table}[h!]
\caption{Summary of transaction delays and costs.}
\begin{adjustbox}{max width=\textwidth}
\begin{tabular}{|l|m{2.3cm}|m{2.9cm}|m{2.3cm}|m{2.9cm}|}
\hline
\textbf{Type of day}
& \textbf{avg. delay of adding entity} 
& \textbf{avg. delay of adding access rule} 
& \textbf{avg. cost for adding entity} 
& \textbf{avg. cost for adding access rule} 
\\ \hline
Busy Day & 0.76 seconds & 17.3 seconds & 0.019 USD & 0.0044 USD  \\ \hline
Average Day & 1.2 seconds & 14.3 seconds & 0.019 USD & 0.0045 USD \\ \hline
Quiet Day & 1.16 seconds & 12.8 seconds & 0.019 USD & 0.0044 USD \\ \hline
\end{tabular}
\end{adjustbox}
\label{tab:Rtable2}
\end{table}

% \textcolor{blue}{Figure~\ref{fig:test1} depicts the distribution of delays for requests made 
% to check the validity of access for a user, when they try to unlock a door using the smart lock. 
% Our results indicate that we can expect a maximum delay of 0.37 seconds to decide whether or not to grant a request.
% This delay is normally acceptable in real applications.}

We now present the distribution of delays obtained 
for verifying user's access privileges, e.g., when meeting participants arrive to the building.
Our simulations indicate that this type of request is completed between 0.26 and 0.37 seconds.
Thus, we can expect a maximum delay of 0.37 seconds to decide whether or not to grant an access request.
This delay is within the acceptable range in real applications.

% \begin{figure}[t]
% \centering
%   \includegraphics[width=\linewidth]{./performance/checkAccess.pdf}
%   \caption[width=.8\linewidth]{Range of response times for check access rule requests.}
%   \label{fig:test1}
% \end{figure}

%-----------------------------------------
\subsection{Scalability}
\begin{figure}[h!]
\centering
  \includegraphics[width=.85\linewidth]{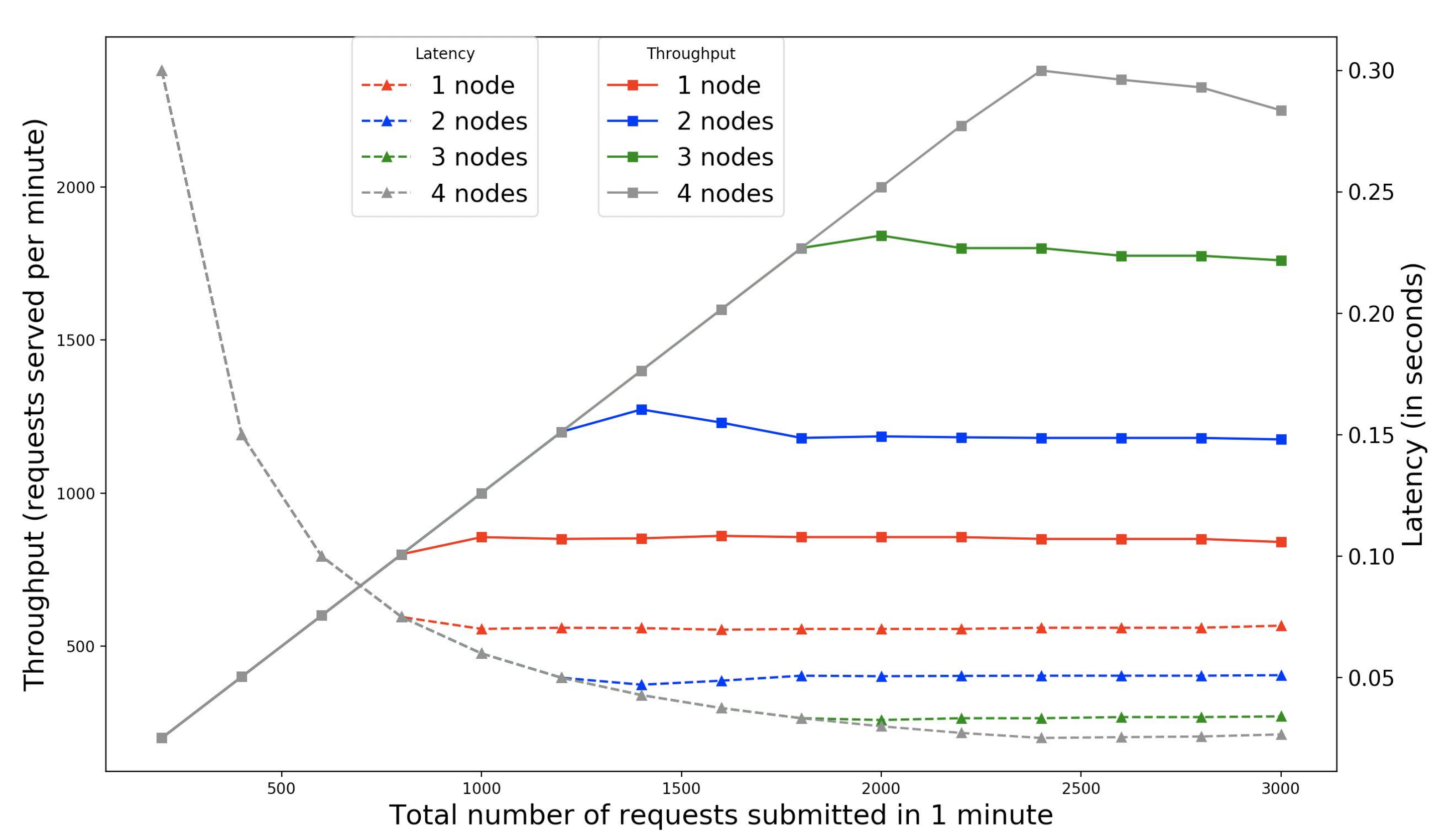}
  \caption{Comparison of throughput and latency for different numbers of nodes and amounts of offered load}
  \label{fig:test2}
\end{figure}

%-----------------------------------------
Lastly, we increase the number of mining nodes in the private blockchain network from 1 to 4.
% Recall that each node represents a group or an organization housed in the commercial building.
We consider two performance metrics, namely throughput and average latency,
and we measure performance as the system load (the total number of transactions) increases in each case.
We assume that transactions are evenly distributed among the nodes in the network.
Figure~\ref{fig:test2} shows the performance evaluation results obtained from 5 independent runs 
for an average day schedule.
It can be seen from the figure that (a) the average latency decreases in most cases as we increase the offered load, 
and (b) throughput increases linearly with the offered load until it reaches a maximum, 
which depends on the number of nodes.
When the load exceeds this threshold, performance starts to fall apart.
Observe that increasing the number of nodes improves performance in terms of both throughput and latency in general.
With 4 nodes in the blockchain network, 
according to Figure~\ref{fig:test2}, we can handle around 2400 requests 
per minute with an average latency as low as 30 milliseconds.

%\begin{figure}[h!]\centering
%\subfloat{\includegraphics[width=.45\linewidth]{./performance/checkAccess.pdf}\caption{Range of response times for check access rule requests.}}\hfill

%\subfloat{\includegraphics[width=.32\linewidth]{./performance/throughput.pdf}  \caption{Comparison of throughput and latency for different nodes and loads}}
%\label{fig}
%\end{figure}

% \begin{figure}[h!]
% \centering
% \begin{minipage}{.45\textwidth}
%   \centering
%   \includegraphics[width=\linewidth]{./performance/checkAccess.pdf}
%   \caption[width=.8\linewidth]{Range of response times for check access rule requests.}
%   \label{fig:test1}
% \end{minipage}
% \begin{minipage}{.45\textwidth}
%   \centering
%   \includegraphics[width=\linewidth]{./performance/throughput.pdf}
%   \caption{Comparison of throughput and latency for different numbers of nodes and amounts of offered load}
%   \label{fig:test2}
% \end{minipage}
% \end{figure}
%-----------------------------------------
\section{Practical Considerations}\label{sec:system}
%-----------------------------------------
\noindent\textbf{Privacy:} Maintaining privacy on blockchain is a complicated issue because transactions and user's balances in a blockchain are open to public viewing. To tackle the privacy issue, Kosba et al.~\cite{kosba2016hawk} build a tool, called `Hawk', which helps developers create privacy-preserving smart contracts without the need of cryptography. The tool is responsible for compiling smart-contract code to a privacy-preserving version. 
Watanabe et al.~\cite{watanabe2015blockchain} propose encrypting smart contracts before deploying them to the blockchain network so only those participants who have the key can access the contract's content (i.e., the state).
Bernable et al.~\cite{8888155} provide a comprehensive review of privacy preserving blockchain approaches.
For example, secure multi-party computation splits the smart contract between 
a number of parties with secret keys to compute parts of the smart contract 
so that a complete picture of a smart contract is not given.
Zero-knowledge proofs can provide verification of smart contracts without 
revealing any information except for the proof to be true; this process can be quite costly.
Commitment schemes allow for proofs to be verified with minimal disclosure of secrets.
Mixing is also an option where transactions are hidden by generating additional transactions 
to create noise and hide the original transaction.
% creating transaction noise to conceal transactions
% Mixcoin: Anonymity for bitcoin with accountable mixes
% https://eprint.iacr.org/2019/341.pdf
% https://eprint.iacr.org/2017/881.pdf
% https://link.springer.com/chapter/10.1007/978-3-319-70278-0_8
% http://zerocash-project.org/media/pdf/zerocash-oakland2014.pdf
Furthermore, user privacy can also be maintained in a hybrid blockchain solution 
where identity is managed by an external public blockchain service, 
while access smart contracts are maintained on a private blockchain.
Our proof-of-concept implementation uses a private Ethereum network which addresses privacy concerns 
to some extent as all participating nodes are within the organization.
Nevertheless, any of the above approaches can be implemented on top of our access-control service 
when the meeting participants and times are sensitive and must be protected from some nodes in the network.
\smallskip

\noindent\textbf{Transaction fees:} Cryptocurrency fees are a fundamental part of blockchain-based software platforms. In some public ledgers there is a minimum fee required for a transaction to be accepted, which helps avoid unwanted and inappropriate transactions. 
In our work, the creation of entities and access rules require a fee, 
which has to be paid by the entity initiating and signing the transaction. 
However, checking access privileges does not cost transaction fees, 
which constitute the most common type of operations. 
When a user tries to access a resource, the API is called to query information from the smart contract and 
decide whether the requested access is allowed. This can be done from any resource and does not incur any fee. 
It should also be noted that with a private Ethereum network, 
% transaction fees are less of a concern as ether can be allocated to the accounts of participating nodes as needed.} 
the difficulty of the mining process can be decided by the management, 
and ether could be mined and potentially transferred to the accounts of participating nodes as needed.
\smallskip

\noindent\textbf{Block time:} Transactions take time to get accepted into the blockchain. However, verifying access privileges does not need to execute transactions to query information from the smart contract's data. The information can be retrieved immediately from the blockchain by the devices when the API is called. 
Apart from this, the creation of new entities and access rules can have long delays in execution. This implies that users may have to wait for some time till their access privileges are granted, since this process requires the execution of two transactions, namely (a) the creation of a new entity and (b) the creation of a new access rule. To mitigate the potentially long wait times, one might raise the transaction fees for the creation and revocation operations to minimize the time spent adding or removing the entities and access rules into the blockchain network~\cite{novo2018blockchain}.
\smallskip

% \noindent\textcolor{red}{\textbf{Roles as groups of paths:} Assigning a group of paths has some limitations that should be considered before implementation. For instance, if roles are used to give permissions, then the access from different people will not be discernible on the blockchain without further secondary evaluation of a person using methods such as a pin code to identify different people. Furthermore, each path may have different temporal restrictions and therefore multiple roles will have to be created anyways to account for the temporal restrictions.}

%copied from novo, should be rephrased
%One possible solution to overcome that disadvantage is the introduction of an expiration date parameter in the operations of the smart contract. This way, the policy rules can expire automatically after a certain time\cite{novo2018blockchain}. This solution does not solve the case in which a rule in a smart contract has to be revoked immediately before its expiration time. Those specific cases are still very hard to solve using the blockchain.

% KE: From meeting, include Architecture diagram, Component diagram, Lab picture
% 1-tier or serverless (time dependent)
%-----------------------------------------
\section{Conclusion and Future Work}\label{sec:conclusion}
%-----------------------------------------
In this paper we proposed a methodology that supports reasoning about and 
flexibly managing the access privileges of occupants and visitors in smart buildings.
Our methodology combines the Brick and BOT models of a building to plan an indoor path 
between two locations and determine the path's ``cost'',
in terms of the sensitivity of the spaces it goes through and the equipment contained in these spaces. 
Furthermore, our method uses smart contracts to manage the space and equipment access privileges of users, 
at specific times and subject to specific constraints.
The unification of RDF graphs that portray building metadata and smart contracts for authorization 
is done through the use of API endpoints to interact with the smart contracts. 
We demonstrated through an example use case that the proposed access-control methodology
is suitable for managing access privileges in large multi-tenant commercial buildings and 
can greatly simplify the labor-intensive security protocol that is currently being followed in such buildings.

%KE: Future Work: Applying a dynamic risk based access system based on blockchain that allows for the overriding of actions in extreme circumstances (e.g. a fireman needing access in case of fire emergency or policeman needing access to investigate a disturbance)

%\subsection{Future Work}
% By integrating two different systems, 
% we have been able to develop applications that are more reliable and manageable for the built environment. 
With small modifications to interact with the actual building's BMS, 
the system can be extended to enable real-world applications without changing the authorizing smart contracts. 
%We also intend to make the creation of resource RDF graphs 
%from using raw building metadata a fully automated process. 
%This can be done by takes advantage of unsupervised learning algorithms. 
Furthermore, incorporating the use of roles instead of individual users receiving access is another extension that we plan to make in future work. Using role-based access control could help group similar types of users into one role and provide them with identical access. For example, all the occupants of an office room with multiple workstations should have access to the same set of physical resources. This approach would be applicable in single-tenant buildings.

Finally, we plan to investigate how to reduce the cost (ether) of using smart contracts.
%We also want to include fail-safe mechanisms in the code to prevent dead-ends. 
A comparative study of using data from RRS and smart contracts directly will be performed in terms of cost and time. 
Apart from authorization, authentication, and revocation, it is possible to build several other applications 
on top of the existing authorizing smart contracts and the building metadata schema, 
thanks to adaptability and flexibility of the proposed solution.

\section*{Appendices}
\appendix

\section{Example Building}
\subsection{Building Graph Model}\label{appendix:graph}
% maybe like http://viewer.brickschema.org/static/soda.pdf
Below is a small sub-graph of the building's model (in RDF syntax) that can be queried using SPARQL as discussed in Section~\ref{sec:queries}.
{\footnotesize
\begin{minted}{turtle}
@prefix bf: <https://brickschema.org/schema/1.0.3/BrickFrame#> .
@prefix bot: <https://w3id.org/bot#> .
@prefix brick: <https://brickschema.org/schema/1.0.3/Brick#> .
@prefix building1: <http://building1.com#> .
@prefix rdf: <http://www.w3.org/1999/02/22-rdf-syntax-ns#> .
@prefix rdfs: <http://www.w3.org/2000/01/rdf-schema#> .
@prefix xml: <http://www.w3.org/XML/1998/namespace> .
@prefix xsd: <http://www.w3.org/2001/XMLSchema#> .

building1:AHU-1 a brick:AHU ;
    bf:feeds brick:VAV-1-12 .

building1:HVAC-Zone-1-12 a brick:HVAC ;
    bf:hasPart building1:Room-1-1-120,
        building1:Room-1-1-121 .

building1:Operations-Zone a bot:Zone ;
    bot:hasSpace building1:Room-1-1-120,
        building1:Room-1-1-121 .

building1:Room-B-100 bf:isLocationOf brick:AHU .

building1:VAV-1-12 a brick:VAV ;
    bf:feeds brick:HVAC-Zone-1-12 ;
    bf:hasPoint building1:Reheat-Command-1-12,
        building1:Temperature-Sensor-1-12,
        building1:Temperature-Setpoint-1-12 .

building1:Temperature-Sensor-1-12 a brick:Temperature_Sensor ;
    bf:controls building1:Reheat-Command-1-12 .

building1:Temperature-Setpoint-1-12 a brick:Temperature_Setpoint ;
    bf:controls building1:Reheat-Command-1-12 .

building1:Door-1-1-12 a bot:Element .

building1:Room-1-1-120 a brick:Room,
        bot:Space ;
    bot:adjacentElement building1:Door-1-1-12 .

building1:Room-1-1-121 a brick:Room,
        bot:Space ;
    bf:isLocationOf brick:Temperature-Sensor-1-12,
        brick:Temperature-Setpoint-1-12 ;
    bot:adjacentElement building1:Door-1-1-12 .

building1:Reheat-Command-1-12 a brick:Heating_Command .
\end{minted}
}
% \subsection{Building Automation System}
% Layout including all room numbers and hvac system, obtained from a screenshot of a building automation system

\subsection{Namespaces for Queries}\label{appendix:namespace}
{\footnotesize
\begin{minted}{sparql}
PREFIX bf: <https://brickschema.org/schema/1.0.3/BrickFrame#> .
PREFIX building1: <http://building1.com#> .
PREFIX brick: <https://brickschema.org/schema/1.0.3/Brick#> .
PREFIX bot: <https://w3id.org/bot#> .
PREFIX rdfs: <http://www.w3.org/2000/01/rdf-schema#> .
\end{minted}
}

\subsection{Building Security Zones}\label{appendix:security_zones}
\begin{figure}[h!]
    \centering
    \includegraphics[width=.8\textwidth]{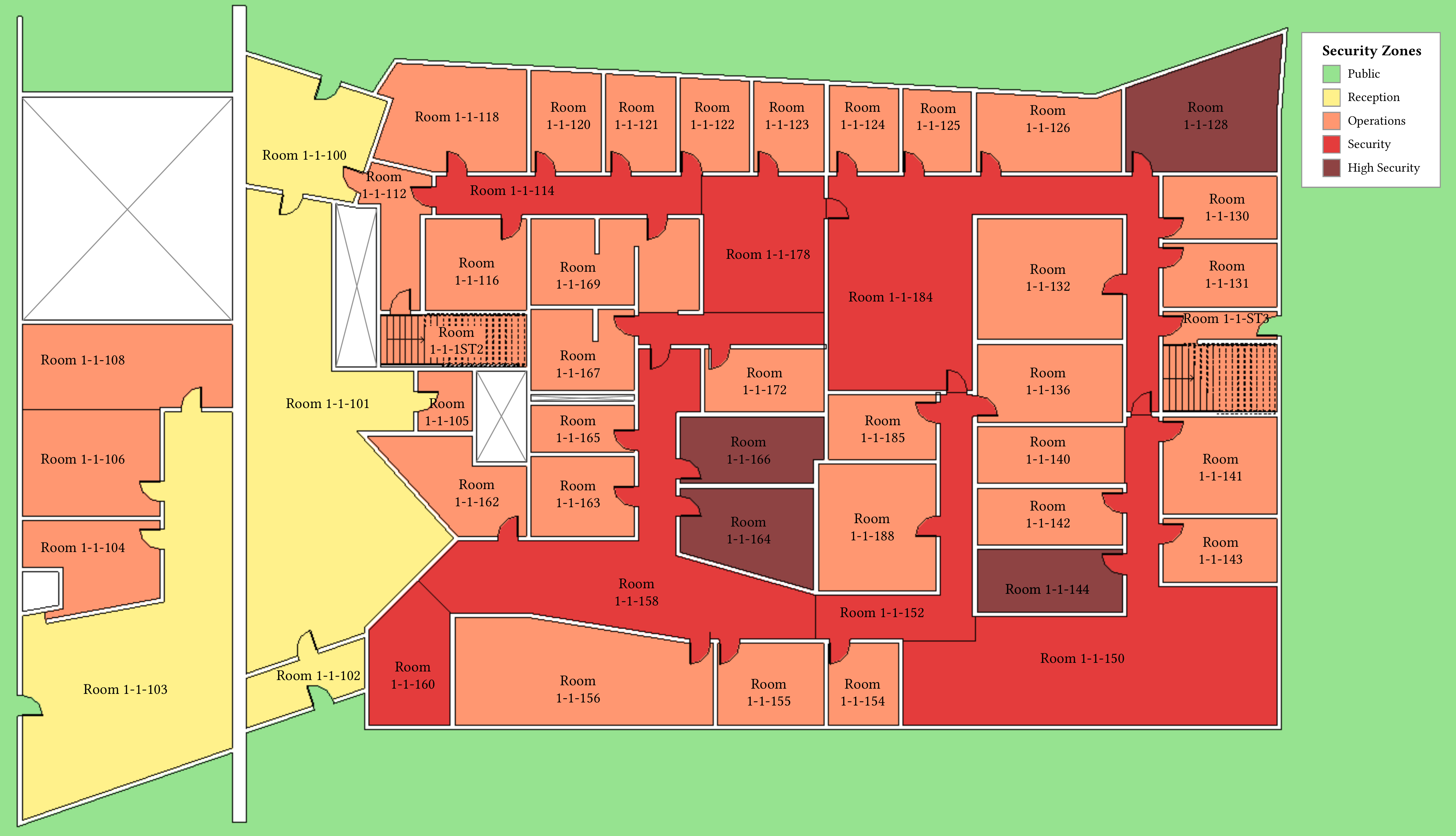}
    \caption{The floor plan of our example building showing the classification of security zones}
    \label{fig:my_label}
\end{figure}

The classification of security zones is carried out by the building manager based on the function of each room in the example building.
Figure~\ref{fig:my_label} illustrates this classification.

\subsection{Example Pathways}\label{pathways}
\subsubsection{Pathway 1}
Below is the cost calculation for \emph{Pathway 1} from Figure~\ref{fig:building} using the cost function defined in Section~\ref{sec:queries}.
{\footnotesize
\begin{align*}
\text{Pathway~1} \equiv \{ &\text{Room\_1-1-102}, \text{Room\_1-1-101}, \text{Room\_1-1-100}, \text{Room\_1-1-112}, \text{Room\_1-1-114}, \\
&\text{Room\_1-1-178}, \text{Room\_1-1-184}, \text{Room\_1-1-152}, \text{Room\_1-1-150}, \text{Room\_1-1-144} \}
\end{align*}
\vspace{-.5cm}
\begin{align*}
\text{cost}(\text{Room\_1-1-102}) &= 1 + 0 \\
\text{cost}(\text{Room\_1-1-101}) &= 1 + 0 \\
\text{cost}(\text{Room\_1-1-100}) &= 1 + 0 \\
\text{cost}(\text{Room\_1-1-112}) &= 3 + 0 \\
\text{cost}(\text{Room\_1-1-114}) &= 3 + 0 \\
\text{cost}(\text{Room\_1-1-178}) &= 3 + 0 \\
\text{cost}(\text{Room\_1-1-184}) &= 3 + ((0.204 \times (1 + 5))+(0.246 \times (1 + 5)) +(0.347 \times (1 + 5)) + (0.413 \times (1 + 5))) \\
\text{cost}(\text{Room\_1-1-152}) &= 3 + 0 \\
\text{cost}(\text{Room\_1-1-150}) &= 3 + (0.413 \times (1+4)) \\
\text{cost}(\text{Room\_1-1-144}) &= 4 + ((0.413 \times (1 + 20))+(0.260 \times (1 + 20)) +(0.328 \times (1 + 20)))\\
\text{cost}(\text{Pathway~1}) &= 55.35
\end{align*}
}

\subsubsection{Pathway 2}
Below is the cost calculation for \emph{Pathway 2} from Figure~\ref{fig:building} using the cost function defined in Section~\ref{sec:queries}.
{\footnotesize
\begin{align*}
\text{Pathway~2} \equiv \{ &\text{Room\_1-1-1ST3}, \text{Room\_1-1-184}, \text{Room\_1-1-150}, \text{Room\_1-1-144} \}\\
\end{align*}
\vspace{-.5cm}
\begin{align*}
\text{cost}(\text{Room\_1-1-1ST3}) &= 2 + 0 \\
\text{cost}(\text{Room\_1-1-184}) &= 3 + ((0.204 \times (1 + 5))+(0.246 \times (1 + 5)) +(0.347 \times (1 + 5)) + (0.413 \times (1 + 5))) \\
\text{cost}(\text{Room\_1-1-150}) &= 3 + (0.413 \times (1+4)) \\
\text{cost}(\text{Room\_1-1-144}) &= 4 + ((0.413 \times (1 + 20))+(0.260 \times (1 + 20)) +(0.328 \times (1 + 20)))\\
\text{cost}(\text{Pathway~2}) &= 42.35 \\
\end{align*}
}

\section{Analytic Hierarchy Process}
\subsection{Questions and Answers}\label{ahpqa}
We ask questions to compare the importance of different types of \emph{Points}, e.g., \emph{sensors} and \emph{setpoints}.
The \emph{sensors} installed in our example building are temperature, damper position, humidity, and occupancy sensors.
The \emph{setpoints} available in our example building are temperature, humidity, and airflow setpoints.
For each set of \emph{Points}, we do a pairwise comparison of its elements using questions, such as \textbf{\emph{How important/critical is element $x$ compared to element $y$ in the building?}}. 
We answer these questions using the rating scale described in Appendix~\ref{ahprating}. 
Each set of questions achieve a consistency ratio of less than $0.10$ as recommended for the AHP process.

For the \emph{sensors}, we achieve a consistency ratio of $0.024$ answering the following questions:
{\footnotesize
\begin{itemize}
    \item How important is Temperature Sensor compared to Damper Sensor? 2
    \item How important is Temperature Sensor compared to Occupancy Sensor? 1
    \item How important is Temperature Sensor compared to Humidity Sensor? 2
    \item How important is Damper Sensor compared to Occupancy Sensor? 1
    \item How important is Damper Sensor compared to Humidity Sensor? 1
    \item How important is Occupancy Sensor compared to Humidity Sensor? 1
\end{itemize}}
For the \emph{setpoints}, we achieve a consistency ratio of $0.055$ answering the following questions:
{\footnotesize
\begin{itemize}
    \item How important is Temperature Setpoint compared to Airflow Setpoint? 2
    \item How important is Temperature Setpoint compared to Humidity Setpoint? 1
    \item How important is Airflow Setpoint compared to Humidity Setpoint? 1
\end{itemize}}
The resulting weights from answering these questions are shown in Table~\ref{AHPtable}.

{\footnotesize
\subsection{Rating Scale}\label{ahprating}
\begin{table}[H]
% \footnotesize
\caption{This rating scale is taken from~\cite{saaty1990analytic} and used to compare 
two elements relative to each other in terms of their importance. 
Note that the reciprocal can also be used to compare elements in the other direction.}
\begin{tabular}{|c|l|}
\hline
\textbf{Rating} & \textbf{Definition}                              \\ \hline
1                                & Equal importance                                 \\ \hline
2                                & Equal to moderate importance of one over another \\ \hline
3                                & Moderate importance                              \\ \hline
4                                & Moderate to essential importance                 \\ \hline
5                                & Essential or strong importance                   \\ \hline
6                                & Essential to very strong importance              \\ \hline
7                                & Very strong importance                           \\ \hline
8                                & Very strong to extreme importance                \\ \hline
9                                & Extreme importance                               \\ \hline
\end{tabular}
\label{ahpratingtable}
\end{table}
}

\bibliographystyle{ACM-Reference-Format}
\bibliography{sigproc}


\begin{thebibliography}{43}

%%% the Web address.

\ifx \showCODEN    \undefined \def \showCODEN     #1{\unskip}     \fi
\ifx \showDOI      \undefined \def \showDOI       #1{#1}\fi
\ifx \showISBNx    \undefined \def \showISBNx     #1{\unskip}     \fi
\ifx \showISBNxiii \undefined \def \showISBNxiii  #1{\unskip}     \fi
\ifx \showISSN     \undefined \def \showISSN      #1{\unskip}     \fi
\ifx \showLCCN     \undefined \def \showLCCN      #1{\unskip}     \fi
\ifx \shownote     \undefined \def \shownote      #1{#1}          \fi
\ifx \showarticletitle \undefined \def \showarticletitle #1{#1}   \fi
\ifx \showURL      \undefined \def \showURL       {\relax}        \fi
% invisible to TeX
\providecommand\bibfield[2]{#2}
\providecommand\bibinfo[2]{#2}
\providecommand\natexlab[1]{#1}
\providecommand\showeprint[2][]{arXiv:#2}

\bibitem[\protect\citeauthoryear{??}{hay}{[n. d.]}]%
        {haystack}
 \bibinfo{year}{[n. d.]}\natexlab{}.
\newblock \bibinfo{title}{Project Haystack}.
\newblock \bibinfo{howpublished}{\url{http://project-haystack.org/}}.
\newblock


\bibitem[\protect\citeauthoryear{Andersen et~al\mbox{.}}{Andersen
  et~al\mbox{.}}{2017}]%
        {Andersen:EECS-2017-234}
\bibfield{author}{\bibinfo{person}{Michael~P Andersen} {et~al\mbox{.}}}
  \bibinfo{year}{2017}\natexlab{}.
\newblock \bibinfo{booktitle}{\emph{{WAVE: A Decentralized Authorization System
  for IoT via Blockchain Smart Contracts}}}.
\newblock \bibinfo{type}{{T}echnical {R}eport} UCB/EECS-2017-234.
  \bibinfo{institution}{UC Berkeley}.
\newblock
\urldef\tempurl%
\url{http://www2.eecs.berkeley.edu/Pubs/TechRpts/2017/EECS-2017-234.html}
\showURL{%
\tempurl}


\bibitem[\protect\citeauthoryear{Andersen et~al\mbox{.}}{Andersen
  et~al\mbox{.}}{2018}]%
        {andersen2018democratizing}
\bibfield{author}{\bibinfo{person}{Michael~P Andersen} {et~al\mbox{.}}}
  \bibinfo{year}{2018}\natexlab{}.
\newblock \showarticletitle{Democratizing authority in the built environment}.
\newblock \bibinfo{journal}{\emph{ACM Transactions on Sensor Networks (TOSN)}}
  \bibinfo{volume}{14}, \bibinfo{number}{3-4} (\bibinfo{year}{2018}),
  \bibinfo{pages}{17}.
\newblock


\bibitem[\protect\citeauthoryear{Andersen et~al\mbox{.}}{Andersen
  et~al\mbox{.}}{2019}]%
        {andersen2019wave}
\bibfield{author}{\bibinfo{person}{Michael~P Andersen} {et~al\mbox{.}}}
  \bibinfo{year}{2019}\natexlab{}.
\newblock \showarticletitle{WAVE: A decentralized authorization framework with
  transitive delegation}. In \bibinfo{booktitle}{\emph{28th {USENIX} Security
  Symposium}}. \bibinfo{publisher}{{USENIX} Association},
  \bibinfo{pages}{1375--1392}.
\newblock


\bibitem[\protect\citeauthoryear{Ardakanian, Bhattacharya, and
  Culler}{Ardakanian et~al\mbox{.}}{2018}]%
        {ardakanian2018}
\bibfield{author}{\bibinfo{person}{Omid Ardakanian}, \bibinfo{person}{Arka
  Bhattacharya}, {and} \bibinfo{person}{David Culler}.}
  \bibinfo{year}{2018}\natexlab{}.
\newblock \showarticletitle{Non-intrusive occupancy monitoring for energy
  conservation in commercial buildings}.
\newblock \bibinfo{journal}{\emph{Energy and Buildings}}  \bibinfo{volume}{179}
  (\bibinfo{year}{2018}), \bibinfo{pages}{311--323}.
\newblock


\bibitem[\protect\citeauthoryear{Balaji et~al\mbox{.}}{Balaji
  et~al\mbox{.}}{2018}]%
        {balaji2016brick}
\bibfield{author}{\bibinfo{person}{Bharathan Balaji} {et~al\mbox{.}}}
  \bibinfo{year}{2018}\natexlab{}.
\newblock \showarticletitle{{Brick: Metadata schema for portable smart building
  applications}}.
\newblock \bibinfo{journal}{\emph{Applied Energy}}  \bibinfo{volume}{226}
  (\bibinfo{year}{2018}), \bibinfo{pages}{1273--1292}.
\newblock


\bibitem[\protect\citeauthoryear{Bazjanac and Crawley}{Bazjanac and
  Crawley}{1999}]%
        {bazjanac1999industry}
\bibfield{author}{\bibinfo{person}{Vladimir Bazjanac} {and} \bibinfo{person}{DB
  Crawley}.} \bibinfo{year}{1999}\natexlab{}.
\newblock \showarticletitle{Industry foundation classes and interoperable
  commercial software in support of design of energy-efficient buildings}. In
  \bibinfo{booktitle}{\emph{Proceedings of Building Simulation}},
  Vol.~\bibinfo{volume}{2}. \bibinfo{pages}{661--667}.
\newblock


\bibitem[\protect\citeauthoryear{{Bernal Bernabe} et~al\mbox{.}}{{Bernal
  Bernabe} et~al\mbox{.}}{2019}]%
        {8888155}
\bibfield{author}{\bibinfo{person}{J. {Bernal Bernabe}} {et~al\mbox{.}}}
  \bibinfo{year}{2019}\natexlab{}.
\newblock \showarticletitle{Privacy-Preserving Solutions for Blockchain: Review
  and Challenges}.
\newblock \bibinfo{journal}{\emph{IEEE Access}}  \bibinfo{volume}{7}
  (\bibinfo{year}{2019}), \bibinfo{pages}{164908--164940}.
\newblock


\bibitem[\protect\citeauthoryear{Bijon, Krishnan, and Sandhu}{Bijon
  et~al\mbox{.}}{2013}]%
        {bijon2013framework}
\bibfield{author}{\bibinfo{person}{Khalid~Zaman Bijon}, \bibinfo{person}{Ram
  Krishnan}, {and} \bibinfo{person}{Ravi Sandhu}.}
  \bibinfo{year}{2013}\natexlab{}.
\newblock \showarticletitle{A framework for risk-aware role based access
  control}. In \bibinfo{booktitle}{\emph{2013 IEEE Conference on Communications
  and Network Security (CNS)}}. IEEE, \bibinfo{pages}{462--469}.
\newblock


\bibitem[\protect\citeauthoryear{Bindra, Lin, Stroulia, and Ardakanian}{Bindra
  et~al\mbox{.}}{2019}]%
        {Bindra19}
\bibfield{author}{\bibinfo{person}{Leepakshi Bindra},
  \bibinfo{person}{Changyuan Lin}, \bibinfo{person}{Eleni Stroulia}, {and}
  \bibinfo{person}{Omid Ardakanian}.} \bibinfo{year}{2019}\natexlab{}.
\newblock \showarticletitle{Decentralized Access Control for Smart Buildings
  Using Metadata and Smart Contracts}. In \bibinfo{booktitle}{\emph{Proceedings
  of the 5th International Workshop on Software Engineering for Smart
  Cyber-Physical Systems}}. \bibinfo{publisher}{IEEE}, \bibinfo{pages}{32--38}.
\newblock


\bibitem[\protect\citeauthoryear{Buterin}{Buterin}{2014}]%
        {buterin2014}
\bibfield{author}{\bibinfo{person}{Vitalik Buterin}.}
  \bibinfo{year}{2014}\natexlab{}.
\newblock \showarticletitle{A next-generation smart contract and decentralized
  application platform}.
\newblock \bibinfo{journal}{\emph{White paper}} (\bibinfo{year}{2014}),
  \bibinfo{pages}{1--37}.
\newblock


\bibitem[\protect\citeauthoryear{Callas et~al\mbox{.}}{Callas
  et~al\mbox{.}}{2007}]%
        {callas2007openpgp}
\bibfield{author}{\bibinfo{person}{Jon Callas} {et~al\mbox{.}}}
  \bibinfo{year}{2007}\natexlab{}.
\newblock \bibinfo{booktitle}{\emph{{OpenPGP} message format}}.
\newblock \bibinfo{type}{{T}echnical {R}eport}.
\newblock


\bibitem[\protect\citeauthoryear{Caronni}{Caronni}{2000}]%
        {caronni2000walking}
\bibfield{author}{\bibinfo{person}{Germano Caronni}.}
  \bibinfo{year}{2000}\natexlab{}.
\newblock \showarticletitle{Walking the web of trust}. In
  \bibinfo{booktitle}{\emph{Proceedings of the 9th International Workshops on
  Enabling Technologies: Infrastructure for Collaborative Enterprises}}. IEEE,
  \bibinfo{pages}{153--158}.
\newblock


\bibitem[\protect\citeauthoryear{Chen and Crampton}{Chen and Crampton}{2011}]%
        {chen2011risk}
\bibfield{author}{\bibinfo{person}{Liang Chen} {and} \bibinfo{person}{Jason
  Crampton}.} \bibinfo{year}{2011}\natexlab{}.
\newblock \showarticletitle{Risk-aware role-based access control}. In
  \bibinfo{booktitle}{\emph{International Workshop on Security and Trust
  Management}}. Springer, \bibinfo{pages}{140--156}.
\newblock


\bibitem[\protect\citeauthoryear{Cruz et~al\mbox{.}}{Cruz
  et~al\mbox{.}}{2018}]%
        {cruz2018rbac}
\bibfield{author}{\bibinfo{person}{Jason Cruz} {et~al\mbox{.}}}
  \bibinfo{year}{2018}\natexlab{}.
\newblock \showarticletitle{{RBAC-SC}: Role-Based Access Control Using Smart
  Contract}.
\newblock \bibinfo{journal}{\emph{IEEE Access}}  \bibinfo{volume}{6}
  (\bibinfo{year}{2018}), \bibinfo{pages}{12240--12251}.
\newblock


\bibitem[\protect\citeauthoryear{Dawson-Haggerty et~al\mbox{.}}{Dawson-Haggerty
  et~al\mbox{.}}{2013}]%
        {Haggerty13}
\bibfield{author}{\bibinfo{person}{Stephen Dawson-Haggerty} {et~al\mbox{.}}}
  \bibinfo{year}{2013}\natexlab{}.
\newblock \showarticletitle{{BOSS}: Building Operating System Services}. In
  \bibinfo{booktitle}{\emph{Proceedings of the 10th {USENIX} Symposium on
  Networked Systems Design and Implementation}}. \bibinfo{publisher}{{USENIX}},
  \bibinfo{pages}{443--457}.
\newblock


\bibitem[\protect\citeauthoryear{Dey}{Dey}{2003}]%
        {doi:10.1061/(ASCE)1527-6988(2003)4:4(213)}
\bibfield{author}{\bibinfo{person}{Prasanta~Kumar Dey}.}
  \bibinfo{year}{2003}\natexlab{}.
\newblock \showarticletitle{Analytic Hierarchy Process Analyzes Risk of
  Operating Cross-Country Petroleum Pipelines in India}.
\newblock \bibinfo{journal}{\emph{Natural Hazards Review}} \bibinfo{volume}{4},
  \bibinfo{number}{4} (\bibinfo{year}{2003}), \bibinfo{pages}{213--221}.
\newblock
\urldef\tempurl%
\url{https://doi.org/10.1061/(ASCE)1527-6988(2003)4:4(213)}
\showDOI{\tempurl}


\bibitem[\protect\citeauthoryear{Dmitrienko et~al\mbox{.}}{Dmitrienko
  et~al\mbox{.}}{2012}]%
        {dmitrienko2012smarttokens}
\bibfield{author}{\bibinfo{person}{Alexandra Dmitrienko} {et~al\mbox{.}}}
  \bibinfo{year}{2012}\natexlab{}.
\newblock \showarticletitle{SmartTokens: Delegable access control with
  {NFC}-enabled smartphones}. In \bibinfo{booktitle}{\emph{International
  Conference on Trust and Trustworthy Computing}}. Springer,
  \bibinfo{pages}{219--238}.
\newblock


\bibitem[\protect\citeauthoryear{Ferraiolo, Cugini, and Kuhn}{Ferraiolo
  et~al\mbox{.}}{1995}]%
        {ferraiolo1995role}
\bibfield{author}{\bibinfo{person}{David Ferraiolo}, \bibinfo{person}{Janet
  Cugini}, {and} \bibinfo{person}{D~Richard Kuhn}.}
  \bibinfo{year}{1995}\natexlab{}.
\newblock \showarticletitle{Role-based access control ({RBAC}): Features and
  motivations}. In \bibinfo{booktitle}{\emph{Proceedings of the 11th annual
  computer security application conference}}. \bibinfo{pages}{241--48}.
\newblock


\bibitem[\protect\citeauthoryear{Hu et~al\mbox{.}}{Hu et~al\mbox{.}}{2014}]%
        {hu2013guide}
\bibfield{author}{\bibinfo{person}{Vincent~C Hu} {et~al\mbox{.}}}
  \bibinfo{year}{2014}\natexlab{}.
\newblock \showarticletitle{Guide to attribute based access control ({ABAC})
  definition and considerations}.
\newblock \bibinfo{journal}{\emph{NIST Special Publication}}
  \bibinfo{volume}{800}, \bibinfo{number}{162} (\bibinfo{year}{2014}).
\newblock


\bibitem[\protect\citeauthoryear{Jacobson et~al\mbox{.}}{Jacobson
  et~al\mbox{.}}{2009}]%
        {jacobson2009networking}
\bibfield{author}{\bibinfo{person}{Van Jacobson} {et~al\mbox{.}}}
  \bibinfo{year}{2009}\natexlab{}.
\newblock \showarticletitle{Networking named content}. In
  \bibinfo{booktitle}{\emph{Proceedings of the 5th International Conference on
  Emerging Networking Experiments and Technologies}}. ACM,
  \bibinfo{pages}{1--12}.
\newblock


\bibitem[\protect\citeauthoryear{Kalam et~al\mbox{.}}{Kalam
  et~al\mbox{.}}{2003}]%
        {kalam2003organization}
\bibfield{author}{\bibinfo{person}{Anas Abou~El Kalam} {et~al\mbox{.}}}
  \bibinfo{year}{2003}\natexlab{}.
\newblock \showarticletitle{Organization based access control}. In
  \bibinfo{booktitle}{\emph{Proceedings of the 4th International Workshop on
  Policies for Distributed Systems and Networks}}. IEEE,
  \bibinfo{pages}{120--131}.
\newblock


\bibitem[\protect\citeauthoryear{Kandala, Sandhu, and Bhamidipati}{Kandala
  et~al\mbox{.}}{2011}]%
        {kandala2011attribute}
\bibfield{author}{\bibinfo{person}{Savith Kandala}, \bibinfo{person}{Ravi
  Sandhu}, {and} \bibinfo{person}{Venkata Bhamidipati}.}
  \bibinfo{year}{2011}\natexlab{}.
\newblock \showarticletitle{An attribute based framework for risk-adaptive
  access control models}. In \bibinfo{booktitle}{\emph{Proceedings of the 6th
  International Conference on Availability, Reliability and Security}}. IEEE,
  \bibinfo{pages}{236--241}.
\newblock


\bibitem[\protect\citeauthoryear{Kosba et~al\mbox{.}}{Kosba
  et~al\mbox{.}}{2016}]%
        {kosba2016hawk}
\bibfield{author}{\bibinfo{person}{Ahmed Kosba} {et~al\mbox{.}}}
  \bibinfo{year}{2016}\natexlab{}.
\newblock \showarticletitle{Hawk: The blockchain model of cryptography and
  privacy-preserving smart contracts}. In \bibinfo{booktitle}{\emph{Symposium
  on Security and Privacy (SP)}}. IEEE, \bibinfo{pages}{839--858}.
\newblock


\bibitem[\protect\citeauthoryear{{Kumar} and {Singh}}{{Kumar} and
  {Singh}}{2016}]%
        {7748957}
\bibfield{author}{\bibinfo{person}{P. {Kumar}} {and} \bibinfo{person}{S.~K.
  {Singh}}.} \bibinfo{year}{2016}\natexlab{}.
\newblock \showarticletitle{{A comprehensive evaluation of aspect-oriented
  software quality (AOSQ) model using analytic hierarchy process (AHP)
  technique}}. In \bibinfo{booktitle}{\emph{Proceedings of the 2nd
  International Conference on Advances in Computing, Communication,
  Automation}}. \bibinfo{pages}{1--7}.
\newblock


\bibitem[\protect\citeauthoryear{Le and Mutka}{Le and Mutka}{2019}]%
        {le2019access}
\bibfield{author}{\bibinfo{person}{Tam Le} {and} \bibinfo{person}{Matt~W
  Mutka}.} \bibinfo{year}{2019}\natexlab{}.
\newblock \showarticletitle{Access control with delegation for smart home
  applications}. In \bibinfo{booktitle}{\emph{Proceedings of the International
  Conference on Internet of Things Design and Implementation (IoTDI)}}. ACM,
  \bibinfo{pages}{142--147}.
\newblock


\bibitem[\protect\citeauthoryear{Narayanan et~al\mbox{.}}{Narayanan
  et~al\mbox{.}}{2016}]%
        {narayanan2016}
\bibfield{author}{\bibinfo{person}{Arvind Narayanan} {et~al\mbox{.}}}
  \bibinfo{year}{2016}\natexlab{}.
\newblock \bibinfo{booktitle}{\emph{Bitcoin and cryptocurrency technologies: a
  comprehensive introduction}}.
\newblock \bibinfo{publisher}{Princeton University Press}.
\newblock


\bibitem[\protect\citeauthoryear{Neuman and Ts'o}{Neuman and Ts'o}{1994}]%
        {neuman1994kerberos}
\bibfield{author}{\bibinfo{person}{B~Clifford Neuman} {and}
  \bibinfo{person}{Theodore Ts'o}.} \bibinfo{year}{1994}\natexlab{}.
\newblock \showarticletitle{Kerberos: An authentication service for computer
  networks}.
\newblock \bibinfo{journal}{\emph{IEEE Communications magazine}}
  \bibinfo{volume}{32}, \bibinfo{number}{9} (\bibinfo{year}{1994}),
  \bibinfo{pages}{33--38}.
\newblock


\bibitem[\protect\citeauthoryear{Novo}{Novo}{2018}]%
        {novo2018blockchain}
\bibfield{author}{\bibinfo{person}{Oscar Novo}.}
  \bibinfo{year}{2018}\natexlab{}.
\newblock \showarticletitle{{Blockchain meets IoT: An architecture for scalable
  access management in IoT}}.
\newblock \bibinfo{journal}{\emph{IEEE Internet of Things Journal}}
  \bibinfo{volume}{5}, \bibinfo{number}{2} (\bibinfo{year}{2018}),
  \bibinfo{pages}{1184--1195}.
\newblock


\bibitem[\protect\citeauthoryear{Ouaddah, Abou~Elkalam, and
  Ait~Ouahman}{Ouaddah et~al\mbox{.}}{2016}]%
        {ouaddah2016fairaccess}
\bibfield{author}{\bibinfo{person}{Aafaf Ouaddah}, \bibinfo{person}{Anas
  Abou~Elkalam}, {and} \bibinfo{person}{Abdellah Ait~Ouahman}.}
  \bibinfo{year}{2016}\natexlab{}.
\newblock \showarticletitle{{FairAccess: a new Blockchain-based access control
  framework for the Internet of Things}}.
\newblock \bibinfo{journal}{\emph{Security and Communication Networks}}
  \bibinfo{volume}{9}, \bibinfo{number}{18} (\bibinfo{year}{2016}),
  \bibinfo{pages}{5943--5964}.
\newblock


\bibitem[\protect\citeauthoryear{Pinno et~al\mbox{.}}{Pinno
  et~al\mbox{.}}{2017}]%
        {pinno2017controlchain}
\bibfield{author}{\bibinfo{person}{Otto Pinno} {et~al\mbox{.}}}
  \bibinfo{year}{2017}\natexlab{}.
\newblock \showarticletitle{Controlchain: Blockchain as a central enabler for
  access control authorizations in the {IoT}}. In
  \bibinfo{booktitle}{\emph{Proceedings of IEEE Global Communications
  Conference (GLOBECOM)}}. IEEE, \bibinfo{pages}{1--6}.
\newblock


\bibitem[\protect\citeauthoryear{Rasmussen et~al\mbox{.}}{Rasmussen
  et~al\mbox{.}}{2017}]%
        {Rasmussen17}
\bibfield{author}{\bibinfo{person}{Mads~Holten Rasmussen} {et~al\mbox{.}}}
  \bibinfo{year}{2017}\natexlab{}.
\newblock \showarticletitle{Recent changes in the building topology ontology}.
  In \bibinfo{booktitle}{\emph{LDAC2017-5th Linked Data in Architecture and
  Construction Workshop}}.
\newblock


\bibitem[\protect\citeauthoryear{Saaty}{Saaty}{1990}]%
        {saaty1990analytic}
\bibfield{author}{\bibinfo{person}{Thomas~L Saaty}.}
  \bibinfo{year}{1990}\natexlab{}.
\newblock \showarticletitle{The Analytic Hierarchy Process}.
\newblock \bibinfo{journal}{\emph{European Journal of Operational Research}}
  \bibinfo{volume}{48} (\bibinfo{year}{1990}), \bibinfo{pages}{9--26}.
\newblock


\bibitem[\protect\citeauthoryear{Saint-Andre}{Saint-Andre}{2005}]%
        {saint2005streaming}
\bibfield{author}{\bibinfo{person}{Peter Saint-Andre}.}
  \bibinfo{year}{2005}\natexlab{}.
\newblock \showarticletitle{{Streaming XML with Jabber/XMPP}}.
\newblock \bibinfo{journal}{\emph{IEEE Internet Computing}}
  \bibinfo{volume}{9}, \bibinfo{number}{5} (\bibinfo{year}{2005}),
  \bibinfo{pages}{82--89}.
\newblock


\bibitem[\protect\citeauthoryear{Salim et~al\mbox{.}}{Salim
  et~al\mbox{.}}{2013}]%
        {salim2013budget}
\bibfield{author}{\bibinfo{person}{Farzad Salim} {et~al\mbox{.}}}
  \bibinfo{year}{2013}\natexlab{}.
\newblock \showarticletitle{Budget-aware role based access control}.
\newblock \bibinfo{journal}{\emph{Computers \& Security}}  \bibinfo{volume}{35}
  (\bibinfo{year}{2013}), \bibinfo{pages}{37--50}.
\newblock


\bibitem[\protect\citeauthoryear{Sandhu et~al\mbox{.}}{Sandhu
  et~al\mbox{.}}{1996}]%
        {sandhu1996role}
\bibfield{author}{\bibinfo{person}{Ravi~S Sandhu} {et~al\mbox{.}}}
  \bibinfo{year}{1996}\natexlab{}.
\newblock \showarticletitle{Role-based access control models}.
\newblock \bibinfo{journal}{\emph{Computer}} \bibinfo{volume}{29},
  \bibinfo{number}{2} (\bibinfo{year}{1996}), \bibinfo{pages}{38--47}.
\newblock


\bibitem[\protect\citeauthoryear{Skandhakumar et~al\mbox{.}}{Skandhakumar
  et~al\mbox{.}}{2016}]%
        {skandhakumar2016graph}
\bibfield{author}{\bibinfo{person}{Nimalaprakasan Skandhakumar}
  {et~al\mbox{.}}} \bibinfo{year}{2016}\natexlab{}.
\newblock \showarticletitle{Graph theory based representation of building
  information models for access control applications}.
\newblock \bibinfo{journal}{\emph{Automation in Construction}}
  \bibinfo{volume}{68} (\bibinfo{year}{2016}), \bibinfo{pages}{44--51}.
\newblock


\bibitem[\protect\citeauthoryear{Skandhakumar et~al\mbox{.}}{Skandhakumar
  et~al\mbox{.}}{2018}]%
        {skandhakumar2018policy}
\bibfield{author}{\bibinfo{person}{Nimalaprakasan Skandhakumar}
  {et~al\mbox{.}}} \bibinfo{year}{2018}\natexlab{}.
\newblock \showarticletitle{A policy model for access control using building
  information models}.
\newblock \bibinfo{journal}{\emph{International Journal of Critical
  Infrastructure Protection}}  \bibinfo{volume}{23} (\bibinfo{year}{2018}),
  \bibinfo{pages}{1--10}.
\newblock


\bibitem[\protect\citeauthoryear{Skandhakumar, Reid, Dawson, Drogemuller, and
  Salim}{Skandhakumar et~al\mbox{.}}{2012}]%
        {skandhakumar2012authorization}
\bibfield{author}{\bibinfo{person}{Nimalaprakasan Skandhakumar},
  \bibinfo{person}{Jason Reid}, \bibinfo{person}{Ed Dawson},
  \bibinfo{person}{Robin Drogemuller}, {and} \bibinfo{person}{Farzad Salim}.}
  \bibinfo{year}{2012}\natexlab{}.
\newblock \showarticletitle{An authorization framework using building
  information models}.
\newblock \bibinfo{journal}{\emph{Comput. J.}} \bibinfo{volume}{55},
  \bibinfo{number}{10} (\bibinfo{year}{2012}), \bibinfo{pages}{1244--1264}.
\newblock


\bibitem[\protect\citeauthoryear{Succar}{Succar}{2009}]%
        {SUCCAR2009357}
\bibfield{author}{\bibinfo{person}{Bilal Succar}.}
  \bibinfo{year}{2009}\natexlab{}.
\newblock \showarticletitle{Building information modelling framework: A
  research and delivery foundation for industry stakeholders}.
\newblock \bibinfo{journal}{\emph{Automation in Construction}}
  \bibinfo{volume}{18}, \bibinfo{number}{3} (\bibinfo{year}{2009}),
  \bibinfo{pages}{357--375}.
\newblock


\bibitem[\protect\citeauthoryear{Wang, Wijesekera, and Jajodia}{Wang
  et~al\mbox{.}}{2004}]%
        {wang2004logic}
\bibfield{author}{\bibinfo{person}{Lingyu Wang}, \bibinfo{person}{Duminda
  Wijesekera}, {and} \bibinfo{person}{Sushil Jajodia}.}
  \bibinfo{year}{2004}\natexlab{}.
\newblock \showarticletitle{A logic-based framework for attribute based access
  control}. In \bibinfo{booktitle}{\emph{Proceedings of the ACM workshop on
  formal methods in security engineering}}. ACM, \bibinfo{pages}{45--55}.
\newblock


\bibitem[\protect\citeauthoryear{Watanabe et~al\mbox{.}}{Watanabe
  et~al\mbox{.}}{2015}]%
        {watanabe2015blockchain}
\bibfield{author}{\bibinfo{person}{Hiroki Watanabe} {et~al\mbox{.}}}
  \bibinfo{year}{2015}\natexlab{}.
\newblock \showarticletitle{Blockchain contract: A complete consensus using
  blockchain}. In \bibinfo{booktitle}{\emph{Proceedings of the 4th global
  conference on consumer electronics (GCCE)}}. IEEE, \bibinfo{pages}{577--578}.
\newblock


\bibitem[\protect\citeauthoryear{Zeilenga}{Zeilenga}{2006}]%
        {zeilenga2006lightweight}
\bibfield{author}{\bibinfo{person}{Kurt Zeilenga}.}
  \bibinfo{year}{2006}\natexlab{}.
\newblock \bibinfo{booktitle}{\emph{Lightweight directory access protocol
  (LDAP): Technical specification road map}}.
\newblock \bibinfo{type}{{T}echnical {R}eport}.
\newblock


\end{thebibliography}
\end{document}